  \input lanlmac.tex
\overfullrule=0pt

\input epsf.tex



\def\figin{\epsfcheck\figin}\def\figins{\epsfcheck\figins}
\def\epsfcheck{\ifx\epsfbox\UnDeFiNeD
\message{(NO epsf.tex, FIGURES WILL BE IGNORED)}
\gdef\figin##1{\vskip2in}\gdef\figins##1{\hskip.5in}
\else\message{(FIGURES WILL BE INCLUDED)}%
\gdef\figin##1{##1}\gdef\figins##1{##1}\fi}
\def\DefWarn#1{}
\def\figinsert{\goodbreak\topinsert}
\def\ifig#1#2#3#4{\DefWarn#1\xdef#1{fig.~\the\figno}
\writedef{#1\leftbracket fig.\noexpand~\the\figno}%
\figinsert\figin{\centerline{\epsfxsize=#3mm \epsfbox{#2}}}
\bigskip\medskip\centerline{\vbox{\baselineskip12pt
\advance\hsize by -1truein\noindent\footnotefont{\sl Fig.~\the\figno:}\sl\ #4}}
\bigskip\endinsert\noindent\global\advance\figno by1}


\font\zfont = cmss10 
\font\litfont = cmr6

\def\bigone{\hbox{1\kern -.23em {\rm l}}}
\def\ZZ{\hbox{\zfont Z\kern-.4emZ}}
\def\hf{{\litfont {1 \over 2}}}

\def\Re{{\rm Re ~}}

\def\p{\partial}

\def\a{\alpha}

\def\g{\gamma}

\def\l{\lambda}




   \def\CD {{\cal D}}
   
   \def\CF {{\cal F}}

   \def\CN {{\cal N}}

   \def\CZ {{\cal Z}}

   \def\R{\relax{\rm I\kern-.18em R}}
   \font\cmss=cmss10 \font\cmsss=cmss10 at 7pt
   \def\Z{\relax\ifmmode\mathchoice
   {\hbox{\cmss Z\kern-.4em Z}}{\hbox{\cmss Z\kern-.4em Z}}
   {\lower.9pt\hbox{\cmsss Z\kern-.4em Z}}
   {\lower1.2pt\hbox{\cmsss Z\kern-.4em Z}}\else{\cmss Z\kern-.4em
   Z}\fi}
   \def\p{\partial}
   
   \def\11{1\!\! 1}

  \def\ran{\right\rangle   }
  \def\lan{\left\langle  }
  \def\({\left(}
\def\){\right)}


\def\diag{{\rm diag}}
\def\Tr{\,{\rm Tr}\,}

\def\Re{{\rm Re}\,}

\def\const{{\rm const}}
\def\tA{\tilde A}
\def\tB{\tilde B}
\def\tC{\tilde C}


\rightline{LPTENS-02/60,\ \ \ \ IHES/P/02/83}
\rightline{FIAN/TD-15/02, ITEP/TH-56/02}
\Title{}
{\vbox{\centerline{ Complex Curve of the Two Matrix Model}
\centerline{ and its Tau-function}
}}
%
%
\centerline{ Vladimir A. Kazakov,$^{1}$\footnote{$^{\ast}$}
{kazakov@lpt.ens.fr} and Andrei Marshakov$^{123}$\footnote{$^{\circ}$}
{{mars@lpi.ru,\ mars@gate.itep.ru}}}

\centerline{$^1${\it  Laboratoire de Physique Th\'eorique de
l'$\acute{E}$cole
Normale Sup\'erieure,\footnote{$^\star$}{{Unit\'e mixte de Recherche
8549 du
Centre National de la Recherche Scientifique et de  l'Ecole Normale
Sup\'erieure et \`a l'Universit\'e de Paris-Sud}} }}
\centerline{{\ \ \ \it 24 rue Lhomond, 75231 Paris CEDEX, France}}
\centerline{$^2${\it Institut des Hautes $\acute{E}$tudes Scientifiques,}}
\centerline{ \it 35 route de Chartres, 91440 Bures-sur-Yvette, France}
\centerline{$^3$ \it Theory Department, P.N.Lebedev Physics Institute and}
\centerline{\it Institute of Theoretical and Experimental Physics,}
\centerline{\it Moscow, Russia\footnote{$^\diamond$}{{Permanent address}}}

\bigskip
%
%
\lref\KKMWZ{I.~K.~Kostov, I.~Krichever, M.~Mineev-Weinstein,
P.~B.~Wiegmann and A.~Zabrodin, ``$\tau$-function for analytic
curves,'' arXiv:hep-th/0005259.}
\lref\KAZ{ V.A.~Kazakov, unpublished; as cited in a footnote of \KKMWZ.}
\lref\MWZ{ M.~Mineev-Weinstein, P.~B.~Wiegmann and A.~Zabrodin,
``Integrable structure of interface dynamics,''
Phys.\ Rev.\ Lett.\ {\bf 84}, 5106 (2000) [arXiv:nlin.si/0001007]. }
\lref\WZ{ P.~B.~Wiegmann and A.~Zabrodin,
``Conformal maps and dispersionless integrable hierarchies,'' Commun.\
Math.\ Phys.\ {\bf 213}, 523 (2000) [arXiv:hep-th/9909147].}
\lref\MaWDVV{
A.~Marshakov, ``On associativity equations,'' Theor.\ Math.\ Phys.\
{\bf 132}, 895 (2002) [Teor.\ Mat.\ Fiz.\ {\bf 132}, 3 (2002)]
arXiv:hep-th/0201267.}
\lref\MaWZ{
A.~Marshakov, P.~Wiegmann and A.~Zabrodin,
``Integrable structure of the Dirichlet boundary problem in two  dimensions,''
Commun.\ Math.\ Phys.\  {\bf 227}, 131 (2002)
[arXiv:hep-th/0109048].}
\lref\KAZBUL{V.A.~Kazakov, ``Ising model on dynamical planar random lattice :
          exact solution'', Phys. Lett.  A 119 (1986) 140;
D.V.~Boulatov, V.A. Kazakov, ``The Ising model on a random planar
          lattice : the structure of phase transition and the exact
          critical exponents'', Phys. Lett. B 186 (1987) 379.}
\lref\DOUGLAS{M.R.~Douglas,  ``The two matrix model'',
 Carg\`ese 1990, Proceedings, Random surfaces and quantum gravity,
77-83.}
\lref\DoWi{R.~Donagi and E.~Witten,
``Supersymmetric Yang-Mills Theory And Integrable Systems,''
Nucl.\ Phys.\ B {\bf 460}, 299 (1996)
[arXiv:hep-th/9510101].}
\lref\Mart{E.~J.~Martinec,
``Integrable Structures in Supersymmetric Gauge and String Theory,''
Phys.\ Lett.\ B {\bf 367}, 91 (1996)
[arXiv:hep-th/9510204].}
\lref\KRICAL{I.Krichever, Func. Anal. \& Appl., {\bf 14} (1980) 282.}
\lref\DKK{ J.-M.~Daul, V.A.~Kazakov, I.K.~Kostov,  ``Rational theories of 2D
                    gravity from the two-matrix model'',
                    Nucl. Phys. B409 (1993) 311.}
\lref\Bou{D.~V.~Boulatov,
``Infinite tension strings at $d > 1$,'' Mod.\ Phys.\ Lett.\ A {\bf 8},
557 (1993), hep-th/9211064.}
\lref\Sta{M.~Staudacher,
``Combinatorial solution of the two matrix model,''
Phys.\ Lett.\ B {\bf 305}, 332 (1993)
[arXiv:hep-th/9301038].}
\lref\Eyn{B.~Eynard,
``Large N expansion of the 2-matrix model,''
arXiv:hep-th/0210047.}
\lref\GM{A.~Gorsky and A.~Marshakov,
``Towards effective topological gauge theories on spectral curves,''
Phys.\ Lett.\ B {\bf 375}, 127 (1996) [arXiv:hep-th/9510224].}
\lref\PQdual{ S.~Kharchev and A.~Marshakov,
``On p - q duality and explicit solutions in $c \leq 1$ 2-d gravity models,''
Int.\ J.\ Mod.\ Phys.\ A {\bf 10}, 1219 (1995) [arXiv:hep-th/9303100].}
\lref\DV{  R.~ Dijkgraaf, C.~ Vafa,
``Matrix Models, Topological Strings, and Supersymmetric Gauge
Theories,'' hep-th/0206255; ``On Geometry and Matrix Models,''
hep-th/0207106; ``A Perturbative Window into Non-Perturbative
Physics,'' hep-th/0208048.}
\lref\Migdal{A.~A.~Migdal,
``Loop Equations And 1/N Expansion,''
Phys.\ Rept.\  {\bf 102}, 199 (1983).}
\lref\SW{ N.Seiberg and E.Witten, ``Monopole Condensation,
And Confinement In N=2 Supersymmetric Yang-Mills Theory'', Nucl. Phys.
{\bf B426} (1994) 19; hep-th/9407087.}
\lref\GKMMM{
A.Gorsky, I.Krichever, A.Marshakov, A.Mironov and A.Morozov,
``Integrability and Seiberg-Witten Exact Solution'', Phys. Lett.  {\bf
B355} (1995) 466; hep-th/9505035. }
\lref\Mbook{
A.~Marshakov, ``Seiberg-Witten Theory And Integrable Systems,'' {\it
Singapore, Singapore: World Scientific (1999) 253 p}. }
\lref\MaPQ{  A.~Marshakov,
``Non-perturbative quantum theories and integrable equations,'' Int.\
J.\ Mod.\ Phys.\ A {\bf 12}, 1607 (1997) arXiv:hep-th/9610242.  }
\lref\sun{
A.Klemm, W.Lerche, S.Theisen and S.Yankielowicz, ``Simple
Singularities and N=2 Supersymmetric Yang-Mills Theory'',
Phys. Lett. {\bf 344B} (1995) 169; hep-th/ 9411048; ``On the
Monodromies of N=2 Supersymmetric Yang-Mills Theory'', hep-th/
9412158;\\ P.Argyres and A.Faraggi, ``The Vacuum Structure and
Spectrum of N=2 Supersymmetric SU(N) Gauge Theory'',
Phys. Rev. Lett. {\bf 73} (1995) 3931, hep-th/9411057.  }
\lref\WDVV{
E.~Witten, ``On the structure of the topological phase of
two-dimensional gravity'', Nucl. Phys. {\bf B340} (1990) 281;
R.~Dijkgraaf, H.~Verlinde and E.~Verlinde, ``Topological strings in
$D<1$'', Nucl. Phys. {\bf B352} (1991) 59.}
\lref\MMM{
A.~Marshakov, A.~Mironov and A.~Morozov, ``WDVV-like equations in N=2
SUSY Yang-Mills Theory'', Phys. Lett. {\bf B389} (1996) 43,
hep-th/9607109. }
\lref\KriW{I.~M.~Krichever,
``The tau function of the universal Whitham hierarchy, matrix models
and topological field theories,'' Commun. Pure. Appl. Math. {\bf 47}
(1992) 437, arXiv:hep-th/9205110.}
\lref\David{ F.~David, ``Non-Perturbative Effects in Matrix Models
and Vacua of Two Dimensional Gravity, Phys.Lett. B302 (1993) 403-410;
hep-th/9212106; F.~David, G.~Bonnet, F.~David, B.~Eynard, ``Breakdown
of universality in multi-cut matrix models,'' J.Phys. {\bf A33} (2000)
6739. }
\lref\KOSTOV{ I.~K.~Kostov, ``Conformal field theory techniques in
random matrix models,'' arXiv:hep-th/9907060.}
\lref\KKN{  V.~A.~Kazakov, I.~K.~Kostov and N.~A.~Nekrasov,
``D-particles, matrix integrals and KP hierarchy,'' Nucl.\ Phys.\ B
{\bf 557}, 413 (1999) hep-th/9810035; J.~Goldstone,
unpublished; J.~Hoppe, MIT PhD thesis, 1982.}
\lref\HKK{  J.~Hoppe, V.~Kazakov and I.~K.~Kostov,
``Dimensionally reduced SYM(4) as solvable matrix quantum mechanics,''
Nucl.\ Phys.\ B {\bf 571}, 479 (2000)
[arXiv:hep-th/9907058].}
\lref\GDKV{ R.~Dijkgraaf, S.~Gukov, V.~A.~Kazakov and C.~Vafa,
``Perturbative analysis of gauged matrix models,''
arXiv:hep-th/0210238. }
\lref\AKK{S.Yu. Aleksandrov, V.A. Kazakov, I.K. Kostov, ``Time-dependent
                    backgrounds of 2D string theory'',
                    Nucl.Phys. B640, (2002) 119-144;
                    hep-th/0205079.}
\lref\IZMehta{C.~Itzykson, J.-B.~Zuber, ``The planar approximation II'',
J. Math. Phys. 21, (1980) 411; M.-L.~Mehta, ``A method of integration
over matrix variables'', Comm. Math. Phys. 79 (1981) 327.}
\lref\GMMMO{
A.~Gerasimov, A.~Marshakov, A.~Mironov, A.~Morozov and A.~Orlov,
``Matrix Models Of 2D Gravity And Toda Theory,'' Nucl.\ Phys.\ B {\bf
357}, 565 (1991).}
\lref\Nekrasov{N.~A.~Nekrasov,
``Seiberg-Witten prepotential from instanton counting,''
arXiv:hep-th/0206161.}
\lref\CheMi{ L.~Chekhov and A.~Mironov,
``Matrix models vs. Seiberg-Witten/Whitham theories,''
hep-th/0209085.}
\lref\KAZMP{V.A.~Kazakov, ``The appearance of matter fields from quantum
          fluctuations of 2D-gravity'', Mod. Phys. Lett. A4 (1989) 2125 }
\lref\DOREY{ N.~Dorey, T.~J.~Hollowood, S.~P.~Kumar and A.~Sinkovics,
``Massive vacua of N = 1* theory and S-duality from matrix models,''
hep-th/0209099. }
\lref\KriUP{I.Krichever, 2000, unpublished. }
\lref\CHUZA{ Ling-Lie Chau, O.~Zaboronsky,
``On the structure of Normal Matrix Model'', Commun.Math.Phys. 196
(1998) 203-247, hep-th/9711091 }
\lref\POTTS{ V.A.~Kazakov, ``Exactly solvable Potts models, bond and tree-like
          percolation on dynamical (random) planar lattice'',
          Nucl. Phys. B 4 (Proc. Supp.) (1988) 93.}
\lref\KAZSM{ V.A.~Kazakov, ``Solvable matrix models'', Random Matrices and their
                    Applications, MSRI publications, volume 40 (2001),
                    hep-th/0003064.}
\lref\KZJ{  V.A.Kazakov, P. Zinn-Justin, ``Two-matrix model with BABA
                    interaction'', preprint hep-th/9808043,
                    Nucl.Phys. B546 (1999) 647-668.}
\lref\KOSADE{ I.K.~Kostov, ``Gauge invariant matrix
model for the A-D-E closed strings,'' Phys.\ Lett.\ B {\bf 297} (1992)
74, hep-th/9208053.}
\lref\COMAMO{
A.~Marshakov, A.~Mironov and A.~Morozov,
``Generalized matrix models as conformal field theories: Discrete case,''
Phys.\ Lett.\ B {\bf 265}, 99 (1991);\ \ \
S.~Kharchev, A.~Marshakov, A.~Mironov, A.~Morozov and S.~Pakuliak,
``Conformal matrix models as an alternative to conventional multimatrix models,''
Nucl.\ Phys.\ B {\bf 404}, 717 (1993)
[arXiv:hep-th/9208044].}
\lref\CDGKV{ F.~Cachazo, R.~Dijkgraaf, S.~Gukov, V.~A.~Kazakov and
C.~Vafa, to appear.}
\lref\ZAW{ P.~Wiegmann, A.~Zabrodin,``Large scale correlations in normal
and general non-Hermitian matrix ensembles'', hep-th/0210159.  }
\lref\AKK{S.Yu.~Aleksandrov, V.A.~Kazakov, I.K.~Kostov, ``Time-dependent
                    backgrounds of 2D string theory'',
                    Nucl.Phys. B640:119-144,2002) 119-144;
                    hep-th/0205079. }
\lref\CDSW{F.~Cachazo, M.~R.~Douglas, N.~Seiberg, E.~Witten,
 ``Chiral Rings and Anomalies in Supersymmetric Gauge Theory'',
 hep-th/0211170.}
\lref\FERRA{ F.~Ferrari, ``Quantum parameter space and double scaling
limits in N=1 super Yang-Mills theory'', hep-th/0211069. }
\lref\VafaR{A.~Klemm, W.~Lerche, P.~Mayr, C.~Vafa, N.~Warner,
``Self-Dual Strings and N=2 Supersymmetric Field Theory'',
Nucl.Phys. B477 (1996) 746-766, hep-th/9604034. }
\lref\HAM{R. Dijkgraaf, M.T. Grisaru, C.S. Lam, C. Vafa, D. Zanon
     `` Perturbative Computation of Glueball Superpotentials'',
hep-th/0211017.}
\lref\AKE{G.~Akemann, ``Higher genus correlators
for the Hermitian matrix models with multiple cuts'', Nucl. Phys.
482, 403 (1996), hep-th/9606004.
}

  \vskip 1cm
\baselineskip8pt{

\baselineskip12pt{
\noindent

We study the hermitian and normal two matrix models in planar
approximation for an arbitrary number of eigenvalue supports. Its
planar graph interpretation is given. The study reveals a general
structure of the underlying analytic complex curve, different from the
hyperelliptic curve of the one matrix model. The matrix model
quantities are expressed through the periods of meromorphic generating
differential on this curve and the partition function of the multiple
support solution, as a function of filling numbers and coefficients of
the matrix potential, is shown to be a quasiclassical
tau-function. The relation to  $\CN=1$ supersymmetric
Yang-Mills theories is discussed. A general class of solvable
multi-matrix models with tree-like interactions is considered. }}

\Date{November, 2002}

\baselineskip=14pt plus 2pt minus 2pt

    \newsec{Introduction}

Among the vast scope of matrix ensembles a distinguished role together
with the integrals over a single matrix is played by two matrix models
- the ensemble of two matrices, usually with the simplest possible
interaction between them \IZMehta. Being still simple integrable
systems (like in the one matrix case their partition function is a
tau-function of Toda lattice hierarchy \GMMMO), these models possess
already a richer mathematical structure than one matrix models and thus
give rise to more applications. The two matrix model was proposed and
studied in \KAZBUL\ as an important solvable example of a new class of
statistical-mechanical systems: spins with nearest neighbor
interaction on planar graphs (Ising spins in this case). Its
multi-critical generalization, in the spirit of the one matrix
multi-critical points \KAZMP, leads to the complete picture of
$(p,q)$-critical points in two-dimensional gravity
\DOUGLAS,\DKK \foot{A similar picture arises in the generalized
one matrix model in external field \PQdual, and it is not surprising
that the multi-support solutions in these models are also related.}.
It appears also in the context of  two-dimensional Laplacian growth
\MWZ,\KKMWZ, \KAZ\ demonstrating some hidden parallels between all
these problems.

Matrix models have been the first physical example when the partition
functions were directly related to tau-functions of integrable systems
\GMMMO. The relation between the (quantum) partition functions and
tau-functions of (classical) integrable systems is still rather
intriguing, it appears to be much more universal than one could
expect before. The same sort of integrable structure like in the
matrix models and/or the topological string theories has been found
\GKMMM\ in the context of Seiberg-Witten theories \SW\ or ${\cal N}=2$
supersymmetric gauge theories in four dimensions.

The similarity between matrix models and supersymmetric gauge theories
based on the similarity of their integrable structures was noticed
long ago \GM.  However, in multidimensional supersymmetric gauge
theories, apart from a recent example \Nekrasov, one mostly observed
the quasiclassical limit of integrable hierarchies (see \Mbook\ for
details and references). It means that the prepotentials of gauge
theories are described rather in terms of the quasiclassical
tau-functions or tau-functions of universal Whitham hierarchy \KriW,
than the tau-functions of dispersionfull hierarchies.

The overwhelming part of the work on the matrix models in the planar
(large $N$) limit concerned the so called one cut case, when the
eigenvalues form a single support distribution, though a few
interesting papers on the multi-support distributions were written in
the past, especially \David\ (see also \KOSTOV), where the relation
with the hyperelliptic curves was revealed. All these papers were
devoted to the one matrix model.

Recently, a new interest to the multi-cut solutions was born, due to
the papers \DV, where the effective superpotentials in the $\CN=1$
supersymmetric gauge theories were related to the matrix models. The
multi-cut solution corresponds there to the breaking of the gauge
group into a few subgroups. It was also proposed in \DV\ to ``fill''
by the eigenvalues not only the minima of the matrix potential, but
also the maxima, the situation missed in the matrix model literature,
due to the obvious absence of stability of such
configurations. However, from the mathematical point of view,
especially from the point of view of the analytic curve description
such filling appears to be admissible and even having some nice
physical applications. These aspects of the multi-cut solution were
further developed in recent papers \GDKV,
\DOREY,\CheMi.

In this paper we are going to study the multi-support solutions in the
two matrix model. As in the one matrix model case, these solutions can
be formulated in terms of geometry of the underlying complex curve
endowed with a generating differential and a quasiclassical
tau-function, of the type proposed in \KriW. We will write down
explicitly the equation for the complex curve in the two matrix model
and define the partition function in the planar limit as a
quasiclassical tau-function. We will also study an important
degenerate case of real potentials and demonstrate the consistency of
the tau-function approach with the planar graph expansion (in terms of
a multi-phase Ising system on the graphs) for the multi-support
solution in the two matrix model. Our method of construction of the
algebraic curve of the two matrix model is powerful enough to be
generalizable to the more general multi-matrix models with the so
called tree-like interactions of matrices.

The two matrix model with the simplest interaction between matrices
is known in the literature in two, superficially different, forms.
The partition function of the normal matrix model (NMM)\foot{ which is
a particular case of the models defined in \CHUZA} of two {\it
commuting} complex conjugated $N\times N$ matrices
$\Phi,\Phi^\dagger$: $[\Phi,\Phi^\dagger]=0$ is defined as follows:
\eqn\NMM{   {\cal Z}_N[t,\bar t\ ]=\int\ \CD \Phi\
\CD \Phi^\dagger\
e^{ -\Tr \Phi\Phi^\dagger +2\Re\Tr W(\Phi)}. }
where the harmonic part of the potential $V(\Phi,\Phi^\dagger) = -
\Phi\Phi^\dagger + 2\Re W(\Phi)$ is parameterized as $W(\Phi)=\sum_{k=1}^K t_k
\Phi^k$. Going to the eigenvalues
$\Phi=\diag(z_1,\ldots,z_N)$ we obtain\foot{ after taking into account
the Jacobian of the angular part of commuting  matrices}:
\eqn\NMMZ{ {\CZ}^{(NMM)}_N[t,\bar t\ ]=
\int\ \prod_{m=1}^N \(d^2z_m\
e^{ - z_m\bar z_m+2\Re W(z_m)}\) |\Delta(z)|^2, }
the normalization by the unitary group volume $V_{U(N)}$ being hidden
into the definition of integration measure in
\NMM. The last integral has a natural interpretation in terms of the
partition function of Coulomb gas of particles with coordinates $z_i$,
confined by the potential in the exponent.

Almost the same eigenvalue integral can be presented as the partition
function of the model of two hermitian {\it non-commuting} matrices
$X,Y$ (H2MM)
\eqn\HMM{ {\cal  Z}^{(H2MM)}_N[t,\tilde t\ ]=\int\ \CD X\ \CD Y\ e^{ -\Tr XY
+\Tr W(X)+ \Tr \tilde W(Y)} }
where $W(X)=\sum_{k=1}^K t_k X^k$ and $\tilde W(Y)=\sum_{k=1}^{\tilde
K} \tilde t_k Y^k$ or, in terms of eigenvalues (using, as usual, the
Harish-Chandra-Itzykson-Zuber (HCIZ) formula):
\eqn\HMMEV{{\cal Z}^{(H2MM)}_N[t,\tilde t\ ]=
\int\ \prod_{m=1}^N \(dx_m\ dy_m \
e^{ - x_my_m + W(x_m)+ \tilde W(y_m)}\) \Delta(x)\Delta(y)}
Now the Jacobian gives $\Delta(x)^2\Delta(y)^2$ in contrast to NMM, but the
extra powers of Vandermonde determinants are canceled by the HCIZ integral.

Indeed, it is not difficult to notice that the H2MM is very similar to
NMM if we take in the former $K=\tilde K$, $\tilde t_k=\bar t_k,\
k=1,2,\ldots$ and compare the eigenvalue representations \NMMZ\ and
\HMMEV. We see that the difference is only in the fact that in \NMMZ\
the eigenvalues of two matrices are complex conjugate, whereas in
\HMM\ they form two independent real sets. The number of the
integration variables is the same, only the contours of integration
are different. It suggests that for a large class of potentials $W$
with good convergence properties at infinity (in particular, for many
polynomial potentials) the partition functions should be equal, and
the correlators of the quantities $\Tr \Phi^n$, $\Tr \Phi^{\dagger m}$
should be the same as correlators of $\Tr X^n$, $\Tr Y^{ m}$. In what
follows we will mostly consider the symmetric case $K = \tilde K$ and
in particular examples even restrict ourselves to the case of real
coefficients $t_k = \bar t_k$.

In the large $N$ limit the models will be also equivalent for most
of the potentials $W$.  The solution to the saddle point equation for H2MM
gives rise in general to  two sets of complex conjugated
eigenvalues: $x_m=z_m,\ \ y_m=\bar z_m\ \ m=1,\ldots,N$, since it is
the only way to make the result for the partition function real for a
general set of complex couplings $t_k$. Hence they will form the same
spots of the two dimensional Coulomb charges with the uniform density
$\rho(z,\bar z)=-\p\bar\p V(z,{\bar z}) = 1$ as in the case of NMM
\WZ, \KKMWZ.

The simplest demonstration of this equivalence comes from the direct
calculation of the Gaussian H2MM integral
\eqn\GAUSS{ \eqalign{
{\cal Z}_N^{Gauss}&=\int \CD X\ \CD Y\ e^{\Tr\left(-XY+ t_1 X+\bar t_1 Y+ t_2
X^2 + \bar t_2 Y^2\right)}   \cr
&=\left({4\pi\over 1-4t_2\bar t_2}\right)^{N^2/2}
\exp N^2\left({\bar t_1t_1+\bar t_1^2t_2
+t_1^2\bar t_2\over 1-4t_2\bar t_2}\right)    }}
which of course coincides with the partition function of NMM with the same
quadratic potential $W(z)$. The latter has in the large $N$
approximation the distribution of the eigenvalues in the shape of
ellipse \WZ, \MaWZ, and the coincidence of results confirms
our statement that
in the saddle point approximation the eigenvalues $x_k, y_k$ will be
complex conjugate and will also form the same ellipse.
Both NMM \KKMWZ\ and H2MM \KAZ\ were proposed as
matrix models describing the two dimensional Laplacian growth processes.

In the next section we will demonstrate the planar diagram technique
for the one and two cut H2MM, relating it to the Ising model and to
the two phase Ising model on planar graphs, respectively. In section 3
we will reproduce (by an unusual method) the solution of the two
matrix model in the planar approximation. Using this solution we will
build in section 4 the general algebraic curve (in general not
hyperelliptic) describing the multi-support two matrix model. We will
describe the topology of its Riemann surface and its possible
degenerations into lower genera. In section 5 the free energy of the
model will be presented as a tau-function in terms of the variables
corresponding to the periods of holomorphic differentials of the
curve. The cubic case with the two cut degeneration will be considered
in detail and a rather explicit solution for its free energy will be
given.  In section 6 we will describe the connection of the two matrix
models (and
of some of their generalizations) to the calculation of effective
superpotentials of $\CN=1$ super Yang-Mills theory with two adjoint
chiral multiplets and an appropriate tree superpotential, in relation
to the conjecture of
\DV. In section 7 we will sketch out the construction of the algebraic
curves for a very general class of solvable matrix models with the
tree-like interactions of the matrices.

\newsec{ Combinatorics of planar graphs of the multi-support two-matrix model}

The equivalence of NMM and H2MM is useful to give a combinatorial
interpretation to the NMM in terms of the planar graph counting.  The
NMM does not have such a direct interpretation but the H2MM does have
it. Indeed, the equivalence between \NMM\ and \HMM\ suggests the
following recipe: if we want to calculate the partition function or
the correlators of traces of normal matrices (without mixture of two
matrices inside each trace!) we simply have to calculate the
corresponding quantities in the corresponding H2MM, taking arbitrary
hermitian matrices $X,Y$ instead of the commuting complex matrices
$\Phi,\Phi^\dagger$, and the same set of complex conjugate couplings
$t_n,\bar t_n$. We can do it by all available methods in the H2MM:
saddle point approximation, orthogonal polynomials or loop equations,
or even perturbatively in the couplings, by the direct planar graph
expansion.

\subsec{ One-support case: Ising spins on planar Feynman graphs }

Let us remind that the H2MM was used in \KAZBUL\ to define and solve
exactly the Ising model on dynamical planar lattices. The role of
these lattices is played by planar Feynman graphs and the positions of
each spin correspond to two types of interaction vertices ($X$-vertex
and $Y$-vertex as spin-up and spin-down). In this respect, we can say
that  the NMM also describes the Ising model on planar graphs. The
phases of the complex couplings correspond here to some generalized
(imaginary) magnetic fields, whose values depend on the phases of
couplings.

For example the two matrix model with the cubic interactions describes
the statistics of Ising spins on $\Phi^3$-type planar graphs (or, due
to the Kramers-Wannier duality, on planar triangulations):
\eqn\CHMM{ {\cal  Z}^{(I)}_N[\g,\l,\bar\l\ ]=
\int\ \CD \Phi\ \CD \Phi^\dagger\ e^{ \Tr\left( -  \Phi^\dagger \Phi
+{\g\over 2}(\Phi^2+ \Phi^{\dagger 2}) + {\l\over 3} \Phi^3 +{\bar\l\over 3}
\Phi^{\dagger 3} \right)} }
It corresponds to the following choice of couplings in \NMM:
$t_1=\bar t_1=0$,
$t_2=\bar t_2=\g/2$, $t_3=\l/3,\bar t_3=\bar\l/3$. Note that this choice
of couplings does not lead to the loss of generality: any cubic
potential can be brought to that used in \CHMM\ by constant shifts of
matrices $X\to X+\const$, $Y\to Y+\overline{\const}$ and the phase
rotations $X\to e^{i\theta}X$, $Y\to e^{-i\theta}Y$.

Let us denote $\l=g\ e^{iH}$.  The planar $\Phi^3$-graphs are
generated in the large $N$ limit by the expansion of \CHMM\ in powers
of $g$, and the corresponding Feynman rules can be given the
statistical mechanical interpretation in terms of  the Ising model on
$\Phi^3$-type planar graphs with the temperature ${2\over
\log\g}$ and the imaginary constant magnetic field $iH$ \KAZBUL.

The solution of two matrix model describing the Ising spins on planar graphs
corresponds to the situation when the eigenvalues for both matrices
form one connected support around the classical minimum of the
potential corresponding to $\Phi=\Phi^\dagger=0$.

\subsec{  Multi-support  case: multi-Ising phases }

 Both NMM and H2MM admit in the large $N$ limit the
multi-support solutions, in analogy with the hermitian one matrix
model, where they were studied from the point of view of their
relation to the hyperelliptic curve in the works \David, \AKE,
\KOSTOV, and recently in \DV\ and \GDKV.
In the case of NMM the eigenvalues $z_k,\bar z_k$ are distributed with
the constant density in a set of disconnected spots on the complex
plane $z$. Our main purpose in this paper is to describe and classify
such solutions from the point of view of the underlying algebraic
curves.

We will work in the notations corresponding to the complex conjugated
$z_k,\bar z_k$ of the NMM. However, all  results will be true for
the independent $z_k,\bar z_k$, as in the H2MM.

The eigenvalue supports appear around the extrema of the potentials.
Note that in the sense of analytic continuation one can also formally
``fill up'' all extrema of the potential and not only the minima. This
leads to more general solutions \DV\ leading to important physical
applications.  The extrema of the potential $V(z,\bar z)=- z\bar z+
W(z)+\overline{W}(\bar z)$ are at the points defined by the system of
equations
\eqn\CLASS{
\bar z = W'(z),\ \ \ \  z = \overline{ W}'(\bar z) }
In general, for the potentials of degree $(n+1)$ we have $n^2$
extrema.

To be more concrete let us study the case of cubic polynomial
potential (related to the one in \CHMM\ by a simple shift of
variables):
\eqn\CUPOTE{
V(z,\bar z)= -z\bar z + T(z+\bar z) +{g\over 3} (z^3+\bar
z^3) }
with real couplings $T$ and $g$, and fill out only the two extrema
obeying the reality condition $z=\bar z$. The classical equations
\CLASS\ for the extrema of \CUPOTE\ can be rewritten in the form of a
"classical" curve
\foot{Compare it  to the curve for the one-support
solution in \Eyn}
\eqn\CUCLA{ \eqalign{
{1\over g^2}\left(\bar z - gz^2 - T\right)\left(z - g{\bar z}^2 - T\right) =
\cr
= z^2{\bar z}^2 - {1\over g}\left(z^3 + {\bar z^3}\right) +
{T\over g}\left(z^2 + {\bar z}^2\right) + {1\over g^2}z{\bar z} -
{T\over g^2}\left(z+{\bar z}\right) + {T^2\over g^2} = 0
}}
and the solution has $n^2=4$ extrema, two extrema for $z={\bar z}$ and
another two for $z+{\bar z} = - 1/g$.  The potential
can be expanded around the extrema $z={\bar z}$ as follows
\eqn\POTSM{ \eqalign{
V(z,\bar z)= -(z-\hat z_a)(\bar z-\hat z_a) + {m_a\over 2} \((z-\hat
z_a)^2+(\bar z-\hat z_a)^2\)+
\cr
 +{g\over 3} \((z-\hat z_a)^3+(\bar z-\hat
z_a)^3\)\pm {\rm const}   }}
where $a=1,2$,
\eqn\CLAZE{
\hat z_{1,2}={1\over2 g} (1\pm\sqrt{1-4T g}) }
and $m_{1,2}=2g\hat z_{1,2}$. We will not consider the filling of the
spots corresponding to other extrema, with $z+{\bar z} = - 1/g$ (later
we will discuss this fact in a more general context).

Let us regroup the eigenvalues into two groups and denote:
\eqn\GROUP{\eqalign{ (z_1-\hat z_1,\ldots,z_{N_1}-\hat z_1)
&= (a_1,\ldots,a_{N_1})\cr (z_{N_1+1}-\hat z_2,\ldots,z_{N}-\hat
z_2)&\equiv (b_1,\ldots,b_{N_2}),\ \ \ N_1+N_2=N }}
 and the same for the conjugated variables
\foot{ In the case of the H2MM one can imagine the situation when the
 filling numbers of the variables $\bar z_i$ are not the same as for
 $z_i$'s: $N'_1\ne N_1,N'_2\ne N_2$; we don't consider this situation
 here.}, corresponding to their positions in the first or second spot,
 respectively. Now we can use the eigenvalue representation \NMMZ\ and
 rewrite this integral in terms of  {\it hermitian} matrices
 $A,\tA$, having the size $N_1\times N_1$ and $B,\tB$ having the size
 $N_2\times N_2$ and a pair of complex rectangular anticommuting ghost
 matrices $C, \tC$ having the size $N_1\times N_2$, as follows:
\eqn\INTM{
{\cal Z}_N[t,\tilde t\ ]= \int\ \CD A\ \CD \tA\ \CD B\ \CD \tB \CD C\
\CD \tC
\ e^{N \Tr S(A,\tA,B,\tB,C,\tC)}     }
where
\eqn\TPOT{ \eqalign{  S(A,\tA,B,\tB,C,\tC)=
&- A\tA + {m_1\over 2} (A^2+\tA^2) + {g\over 3}(A^3+\tA^3)\cr &-B\tB +
{m_2\over 2}(B^2+\tB^2) + {g\over 3}(B^3+\tB^3)\cr &-m C^\dagger C - m
\tC^\dagger \tC - g C^\dagger C A+ g C C^\dagger B-
g \tC^\dagger \tC \tA+ g\tC \tC^\dagger \tB }}
and $m= \sqrt{1-4Tg}$. To rewrite the eigenvalue integrals as the
matrix ones, we used the HCIZ formula for the $U(N_{1,2})$ group
integral
$$
\int [d\Omega]_{U(N_1)}\ e^{\Tr(\Omega^\dagger x \Omega \bar x)} \propto
{\det_{ij}e^{a_i\bar a_j}\over \Delta(a)\Delta(\bar a)}
$$
and similarly for $b,\bar b$.  The matrices $C,\tC$ served to
exponentiate the cross-products $\prod_{k,m}(a_k-b_m)(\bar a_k -\bar
b_m)$ in the Vandermonde determinants in \NMMZ\ (see the similar method
for the one matrix model in \David,\GDKV).

\ifig\planar{ 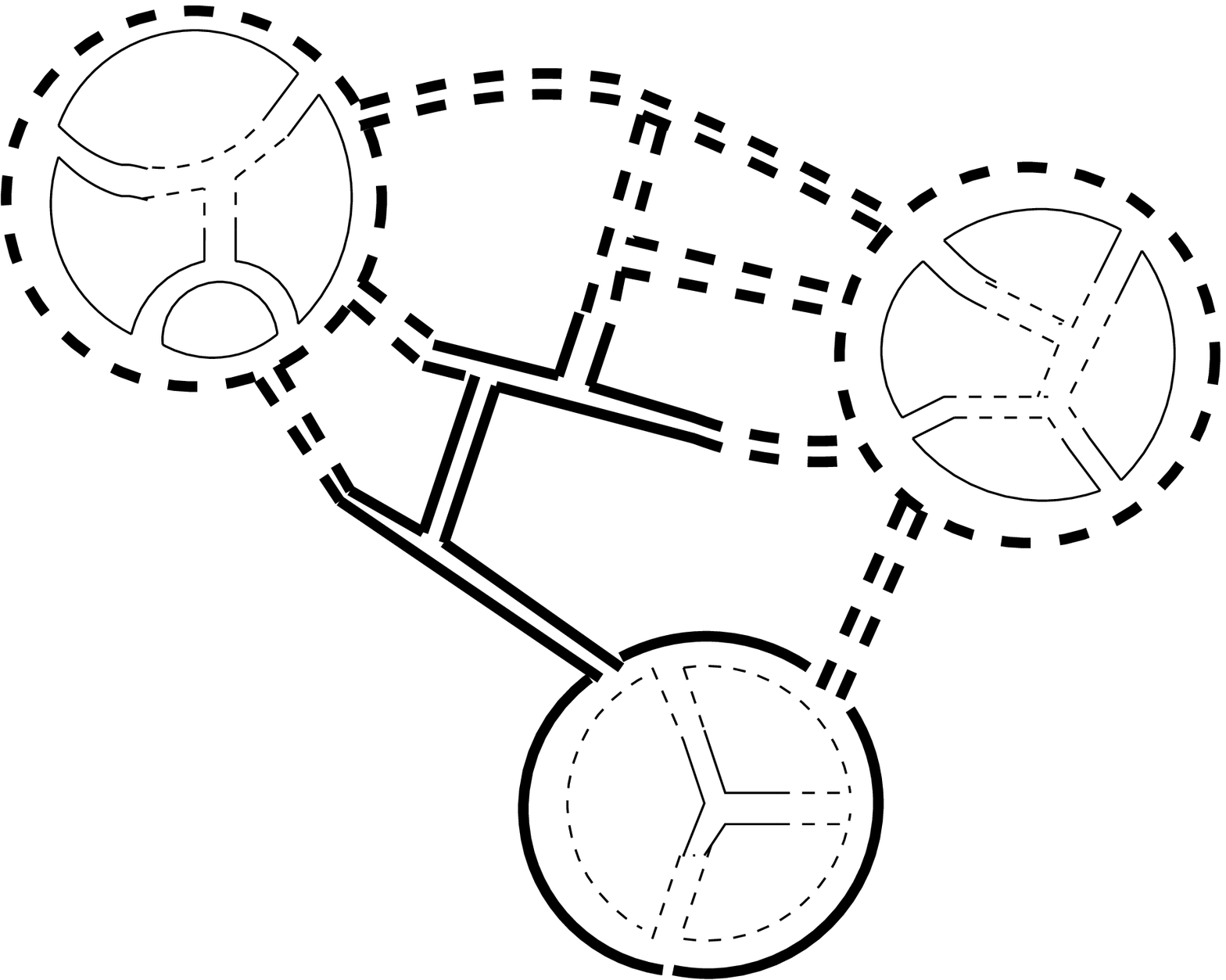}{70}{ A planar graph of the  two matrix model
 with two eigenvalue supports. There are two phases here: thin line
 phase (inside the circles) and thick line phase (outside the
 circles).  Each phase corresponds to two different kinds of Ising
 spins having a different temperature: the spins looking ``up'' are
 located in the triple vertices made of solid double lines, and the
 spins looking ``down'' are located in the triple vertices made of
 dotted double lines. The three types of propagators inside each phase
 (solid, dotted and mixed) describe the interactions depending on the
 mutual orientation of the neighboring spins. Along the interphase
 (ghost) lines drown by circles, the spins have the same orientation.}

Now we can give the model \TPOT\ a combinatorial interpretation in
terms of the planar graph expansion. Namely, we expand \INTM\ in the
cubic coupling $g$, for the fixed $m_1,m_2,m$.  The diagram technique
consists of the following elements:
\eqn\FEYNP{ \eqalign{
{\rm propagators:}\ \ \ &\lan AA\ran_0=\lan \tA\tA\ran_0= {2m_1 \over
m_1^2-1},\
\ \lan A\tA\ran_0= {2\over m_1^2-1}; \cr
&\lan BB\ran_0=\lan \tB\tB\ran_0= {2m_2 \over m_2^2-1},\
\ \lan B\tB\ran_0= {2\over m_2^2-1}\cr
& \lan C^\dagger C\ran_0=\lan \tC^\dagger \tC\ran_0= 1/m  }}
\eqn\FEYNP{ \eqalign{
{\rm vertices:}\ \ \ &\lan AAA\ran_0=\lan \tA\tA\tA\ran_0= \lan
BBB\ran_0=\lan
\tB\tB\tB\ran_0  \cr
&=-\lan C^\dagger C A\ran_0=\lan C C^\dagger B \ran_0 =
-\lan \tC^\dagger \tC \tA \ran_0 =\lan \tC \tC^\dagger \tB \ran_0 =g }}
Each type of the propagators $\lan C^\dagger C\ran_0$ and $\lan
\tC^\dagger \tC\ran_0$ forms closed loops on  Feynman graphs,
each loop entering with the factor $(-1)$.

A typical planar graph for the two-support model is presented on
\planar.  Let us classify the index loops of each planar graph (or a
graph of a fixed topology) as carrying the index $i=1,\ldots,N_1$
(solid line), or the index $i'=1,\ldots,N_2$ (dotted line). The ghost
loops (drawn by a double line formed by a thick and a thin line) will
separate two phases on the planar graph: one described by the matrices
$A,\tA$ (thick line phase) and another described by the matrices
$B,\tB$ (thin line phase). Each of these phases corresponds to the
dynamics Ising spins of 1-st and 2-nd kind, as described in the
previous subsection for the single support case. At the phase
boundaries formed by the ghost loops, the spins have the same
orientation (two types of ghosts $C$ and $\tC$ correspond to two
possible orientations). Each solid index loop contributes a factor
$N_1$, and each dotted index loop -- a factor $N_2$.

Let us note at this point that for $gT<1/4$ we always have real
$m_1>0,\ m_2>0$, but the determinants of second derivatives of the
action at two different extrema are $m_1^2-1>0$ and $m_2^2-1<0$
correspondingly, which means that the first extremum is the true
minimum, and the second is a saddle point of the potential.

The last comment about this diagram technique: as it was done in
\GDKV\ for the two-cut   one matrix model, we can do the
following formal operation with each graph, without changing its
contribution: we can change the contribution of each ghost loop from
$(-1)$ to $1$, change the sign of each $BB$, $\tB\tB$ and $B\tB$
propagator, and the sign of $N_2$ \foot{we will see in the next sections
that $g N_1/N$ and $g N_2/N$ can be viewed as independent
variables in the planar limit, as in the one matrix model case \GDKV}.

All this means that we can consider instead of the matrix model
\INTM-\TPOT, the matrix model with the action:
\eqn\TPOTS{ \eqalign{  S(A,\tA,B,\tB,C,\tC)=
&- A\tA +{ m_1\over 2} (A^2+\tA^2) + {g\over 3}(A^3+\tA^3)\cr &+B\tB -
{m_2\over 2}(B^2+\tB^2) + {g\over 3}(B^3+\tB^3)\cr &-m C^\dagger C - m
\tC^\dagger \tC + g C^\dagger C A+ g C C^\dagger B+
g \tC^\dagger \tC \tA+ g\tC \tC^\dagger \tB  }}
Here $C$ and $\tC$ are already the usual commuting complex $N_1\times
N_2$ rectangular matrices (the sign of $N_2$ is again normal here).
We also changed the variables as follows $A\to -A$, $\tA\to -\tA$.

In this new representation of the same model, the perturbative
$g$-expansion goes around the true minima of the potential, and the
contributions of planar graphs are positive. The planar expansion of
this matrix model define the statistical mechanical model on random
dynamical graphs describing a two phase system, each phase
corresponding to the system of Ising spins with the ferromagnetic
boundary condition on the phase boundary.  Note that since we have two
independent ``cosmological constants'': the coupling $g$ and $N_2/N$,
but only one parameter $T$ related to the (different) temperatures of
two kinds of Ising spins. Hence we cannot make both types of the Ising
spins critical at the same time. We need for that higher powers of the
potential.  In a sense, our multi-cut solution generalizes the ADE
models proposed in \KOSADE,\COMAMO.

Let us conclude this section by noticing that much of what we did here
on the two support case can be carried over to the 4-support case of
this model and to the multiple supports for the potentials of higher
degree. However, unlike the cubic case with real couplings, the
details are difficult to work out. Below we will return to the cubic
potential and discuss in detail the generic 4-support structure.  We
will also see that the generic 4-support solution has a very natural
2-support "degeneration", corresponding precisely to the perturbation
theory considered in this section.

\newsec{ Solution of the model in the planar limit}

Let us now turn to the solution of the two matrix model in the planar
or large $N$ limit. As is well-known in this case the computation of
matrix integrals \NMM\ or
\HMM\ can be reduced to the solution of the saddle point equation.

The saddle point equation for the model \NMMZ\ with the eigenvalues
$z_1,\ldots,z_N$ (or analytically continued saddle point equation for
\HMMEV) reads
\eqn\SPED{  \bar z_k=  W'(z_k)+\sum_{j(\ne k)}{1\over z_k-z_j}  }
together with the complex conjugated equation.  For the resolvents of
distributions of the eigenvalues
\eqn\RES{
G(z)=\hbar\left< \Tr {1\over z-\Phi}\right> ,\ \ \ \bar
G(\bar z)=\hbar\left<\Tr {1\over \bar z-\Phi^\dagger}\right> }
it can be written as:
\eqn\SPE{ \eqalign{  \bar z&= W'(z)+G(z) \cr
z&= \overline{ W}'(\bar z)+\bar G(\bar z) }}
where the resolvent has the usual asymptotics at large $z$ or $\bar z$
for the finite supports:
\eqn\ASSMP{G(z)\to t_0/z+O(1/z^2),\ \ \ \bar
G(\bar z)\to t_0/\bar z+O(1/\bar z^2)}
To fix the resolvents in \SPE we have to impose the condition that the
functions $\bar z(z)$ and $z(\bar z)$ are mutually inverse:
\eqn\INV{    \bar z(z(x))=x }
To justify this condition we recall once again that the solutions of
these equations describe the  spots of Coulomb charges with the uniform
distribution of the eigenvalues with coordinates $(z_1,\bar
z_1),\ldots,(z_N,\bar z_N)$ with the density $\rho(z,\bar z)=1$. The
boundaries of the spots are in general smooth curves in the complex
plane $z$ depending on the couplings of the potential.  To fix the
form of these boundaries it is enough to consider the eqs. \SPE\ at
the boundary. Then both equations should define the same curve $\bar
z(z)$. It means that the solutions of these equations, $\bar z(z)$ and
$z(\bar z)$ respectively, should be mutually inverse, i.e.  obey
eq. \INV
\foot{A more rigorous derivation of these equations from the method of
orthogonal polynomials can be found in \DOUGLAS, \DKK.}.  Note that in
general ${\bar z}$ should be treated as an independent function on the
complex manifold (with involution) and it becomes literally complex
conjugated to the function $z$ only on some real section -- the real
analytic curve in the sense of \KKMWZ, which is just a boundary
of the eigenvalue distribution. To avoid further
misunderstanding in what follows we will denote this function as
${\tilde z}(z)$, so that ${\tilde z}(z) = \bar z$ (i.e. is literally
complex conjugated only on the boundaries of the spots).

In the quasiclassical, or dispersionless limit one considers the free
energy of the matrix ensembles to be defined as a "planar" limit
\eqn\planar{
{\cal F}(t,S) = {\rm lim}\left( \hbar^2\log{\cal
Z}\left({t\over\hbar}\right)\right) }
implying $N\to\infty$, $\hbar\to 0$ with $N\hbar = t_0$ being fixed. In
\planar\ $t$ denote the parameters of the potential $V(z,{\bar z}) =
-z{\bar z} + W(z) + {\overline W}({\bar z})$ while $S$ are the new
variables directly related to the "filling numbers" of various
eigenvalue supports. More strictly, by the planar limit \planar\ one
usually understands the solution to the variational problem
\eqn\variF{ \eqalign{
{\cal F} \propto \int V(z,{\bar z})\rho (z,{\bar z})d^2z -
\int d^2z_1d^2z_2 \rho (z_1,{\bar z}_1)
\rho (z_2,{\bar z}_2)\log\left|z_1-z_2\right| +
\cr
+ \sum_\alpha v_\alpha
\left(\int \rho (z,{\bar z})d^2z - S_\alpha\right) }}
which is the "stationary phase" condition for corresponding matrix
integrals \NMMZ, \HMMEV. Note that the normalization of the density at
different supports is achieved by the Langange multipliers $v_\a$. The
dispersionless tau-function can be obtained from \variF\ by the
substitution of the saddle point solution $\rho = \rho_c$ of the
saddle point equation
$$
{\delta {\cal F}\over\delta \rho(z,{\bar z})} = 0,
$$
or
\eqn\vareq{
v_\alpha =
 \int d^2z' \rho (z',{\bar z}')
\log\left|z_\alpha-z'\right| - V(z_\alpha,{\bar z}_\alpha) }
for any point $P_\alpha = (z_\alpha,{\bar z}_\alpha)$ belonging to one
of the supports, labelled by $\alpha$.

In the next two sections we will first discuss the structure of the
complex curve \SPE\ for two matrix model and then define the free
energy \variF\ as a (logarithm) of quasiclassical tau-function.

\newsec{Complex curve for the two-matrix model}

The reality condition  suggests the following ansatz for
the solution to
\SPE:
\eqn\POLEQ{  F(z,\tilde z)= \sum_{i,j} f_{ij} z^i \tilde z^j =0, }
with the coefficients obeying the symmetry: $f_{ij}=\bar f_{ji}$
\foot{since any such curve should be consistent with its real
section, the eq.  $F(x+iy,x-iy)= {\cal P}(x,y) = 0$ with real
coefficients.}.  Due to this symmetry the equation \INV\ will be
automatically satisfied.

The coefficients  $f_{ij}$ can be partially fixed by the asymptotics:
$$
\left.\tilde z(z)\right|_{z\to\infty}\simeq W'(z)+{t_0\over z}+
O\left({1\over z^2}\right)
$$
following from \SPE\ and \ASSMP, but the number of parameters of the
potential grows linearly with its degree while the number of
coefficients of \POLEQ\ grows quadratically. The rest of the
parameters will correspond to the eigenvalue filling numbers for
various spots (supports of the eigenvalues on the $z$
plane). Altogether they will play the role of moduli of complex
structure of the algebraic curve defined by the eq. \POLEQ.

One may also think of the analytic curve \POLEQ\ as of the algebraic
form of the large $N$ loop equations in H2MM (but not in NMM!) for the
resolvent $G(z)$ \Sta,\Eyn, or for the matrix model in external field
\Bou, where it was first proposed.

\subsec{Structure of the curve}

We can precise the algebraic equation of the curve \POLEQ\ for the
mutually complex conjugated potentials of a degree $K=n+1$ (with a few
explicitly given highest degree terms)
\eqn\complcu{
F(z, {\tilde z}) = z^n{\tilde z}^n + az^{n+1} + {\bar a}{\tilde
z}^{n+1} + \sum_{i,j\in (N.P.)_+} f_{ij}z^i{\tilde z}^j = 0 }
%
\ifig\newton{ 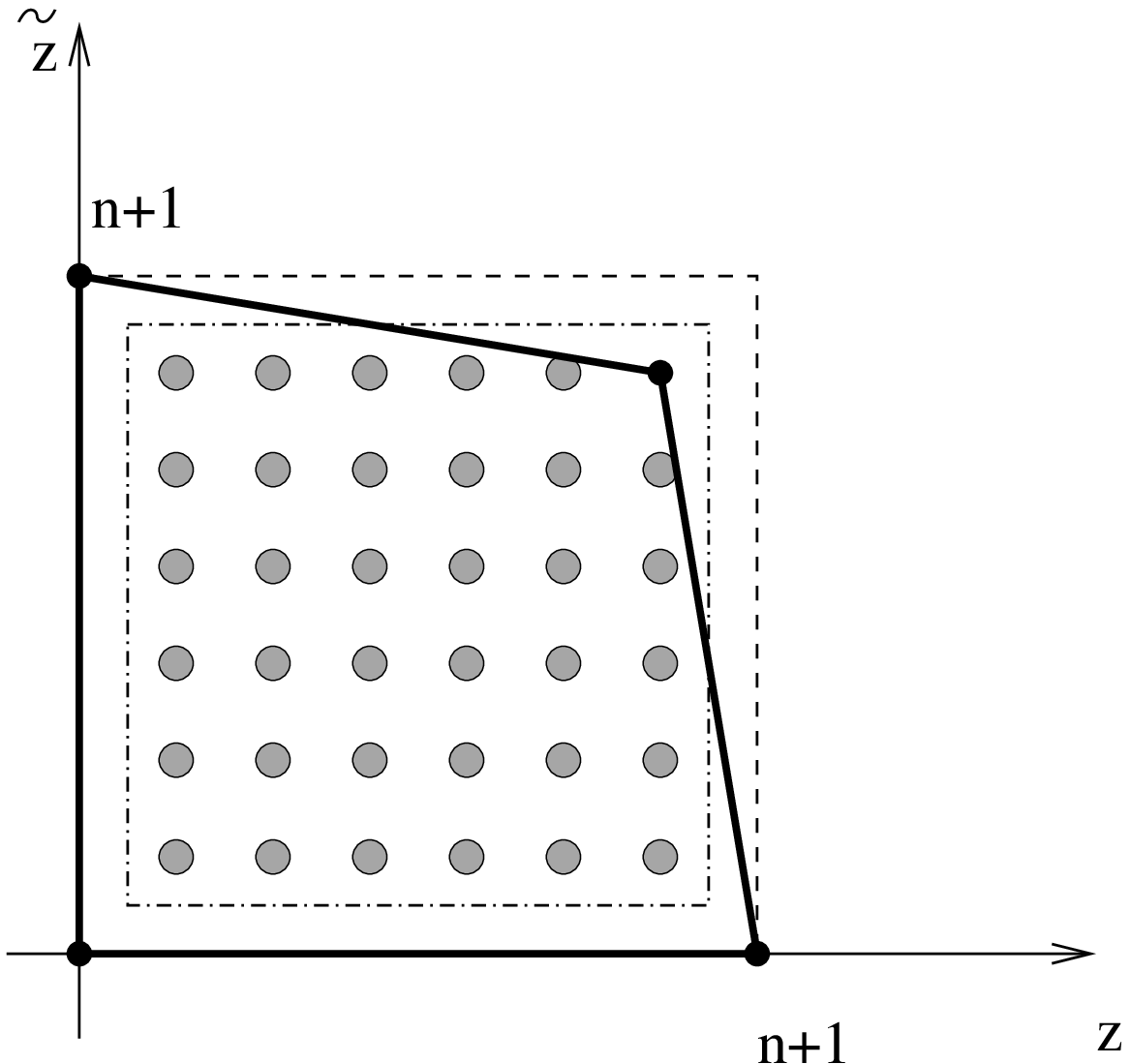}{70}{ The Newton polygon for the curve \complcu.
The highest degree terms in (\complcu\ determine the shape of the
polygon and the integer dots inside it count the number of holomorphic
differentials, or genus of the curve. Clearly this number is equal to
the area of "dual" square except for one (black) point, so that
$g=n^2-1$.}
where the first three terms correspond to the three points on the
boundary lines of the Newton polygon (square in this case) on \newton,
and the sum over $(N.P.)_+$ in the last term means the sum over the
points inside the Newton polygon (including the points on both axis
not marked on \newton). For example, there are 8 terms in this sum for
$n=2$.

One may compare this equation with \CUCLA\ and see that the higher
degree terms are always fixed by the "classical" equations on extrema
of the matrix model potential.  The properties of the curve (\complcu)
can always be easily established via the Newton polygon on \newton.

Counting the number of integer points inside the polygon one finds
that the number of holomorphic differentials, or genus of the curve,
is equal to
\eqn\genus{ g = n^2-1 . }
A simple basis for the holomorphic differentials can be chosen as
\eqn\vij{   dv_{ij} = z^i{\tilde z}^j {d{\tilde z}\over F_z} =
-z^i{\tilde z}^j {dz\over F_{\tilde z}},  }
with the degrees $i=i'-1$ and $j=j'-1$, where $(i',j')\in N.P.$ are
coordinates of the points {\it strictly} inside the Newton polygon, without
the boundary points (see
\newton)
\foot{For example, for $n=2$ there are three points inside the
polygon: $i',j'>0$ and $i'+j'\leq 2$, then the holomorphic differentials
are labeled by $i,j\geq 0$ and $i+j\leq 1$.}.

Finally let us point out here that for the models with non-symmetric
potentials $W$ and $\tilde W$ one may write the spectral curve
equation in a similar way, but it would not obey such symmetric
properties. For the potentials $W$ and $\tilde W$ of degrees $n+1$ and
$\tilde n+1$, correspondingly, the highest terms will always be of a
particular form (one may equally use here $x$ and $y$ instead of $z$ and
$\tilde z$,
to show its direct relation to \HMMEV)
\eqn\NONSY{
F(z,{\tilde z}) = z^n{\tilde z}^{\tilde n} + Az^{n+1} + B{\tilde z}^{\tilde n+1} +
\sum_{(i,j)\in (N.P.)_+} f_{ij}z^i{\tilde z}^j }
and the genus of the curve \NONSY\ is $n{\tilde n}-1$. It means that
the $(n+1)\times(n+1)$ square on \newton\ should be replaced by the
rectangle of the size $(n+1)\times({\tilde n}+1)$ with all other
elements of the construction remaining intact. Of course, for ${\tilde
n}=1$ integrating over the matrix with Gaussian potential one returns
to the 1MM with the (hyperelliptic) curve of genus $n-1$ (see section $5.1$
below).

\subsec{The cubic example}

To understand better the structure of the curve let us first discuss
in detail the cubic example. Writing eq.~\complcu\ first with
arbitrary coefficients
\eqn\cuthree{
F(z,{\tilde z})= z^2{\tilde z}^2 + az^3 + {\bar a}{\tilde z}^3 +
bz^2{\tilde z} + {\bar b}z{\tilde z}^2 + cz^2 + {\bar c}{\tilde z}^2
+ fz{\tilde z} + qz + {\bar q}{\tilde z} + h = 0}
one finds that its structure is in fact very similar to the equation
of the "classical" curve \CUCLA. Indeed, the eq. \cuthree\ should be
consistent with the asymptotics
\eqn\asyone{
{\tilde z} = W'(z) + G(z) = \sum_{k=1}^{3}kt_kz^{k-1} +
O\left({1\over z}\right) \equiv \lambda z^2 + \gamma z + \eta +
O\left({1\over z}\right)}
Substituting \asyone\ into the eq. \cuthree\ and collecting the
coefficients in front of the terms $z^6$, $z^5$ and $z^4$ one gets
\eqn\coeff{ \eqalign{
a = - {1\over\bar\lambda} = -{1\over g}
\cr
b = {\bar\gamma\over\bar\lambda} = 0
\cr
c =  {\bar\eta\over\bar\lambda} -
{\gamma\over\lambda\bar\lambda} + 2{\bar\g^2\over\bar\l^2}= {T\over g}
}}
and their complex conjugated counterparts, i.e. the coefficients at
higher degree terms are indeed completely fixed by parameters of the
potential \CUPOTE. Four lower degree coefficients $f$, $q$, ${\bar q}$
and $h$ correspond to the bipole differential and three holomorphic
differentials~\foot{Note, that the curve \complcu,\ \cuthree\ is
written implying some reality condition onto the coefficients, but as
usual, the deformations of these coefficients should be considered as
independent complex variables.}. Their classical "expectation values"
are presented in eq. \CUCLA.

\ifig\cutre{ 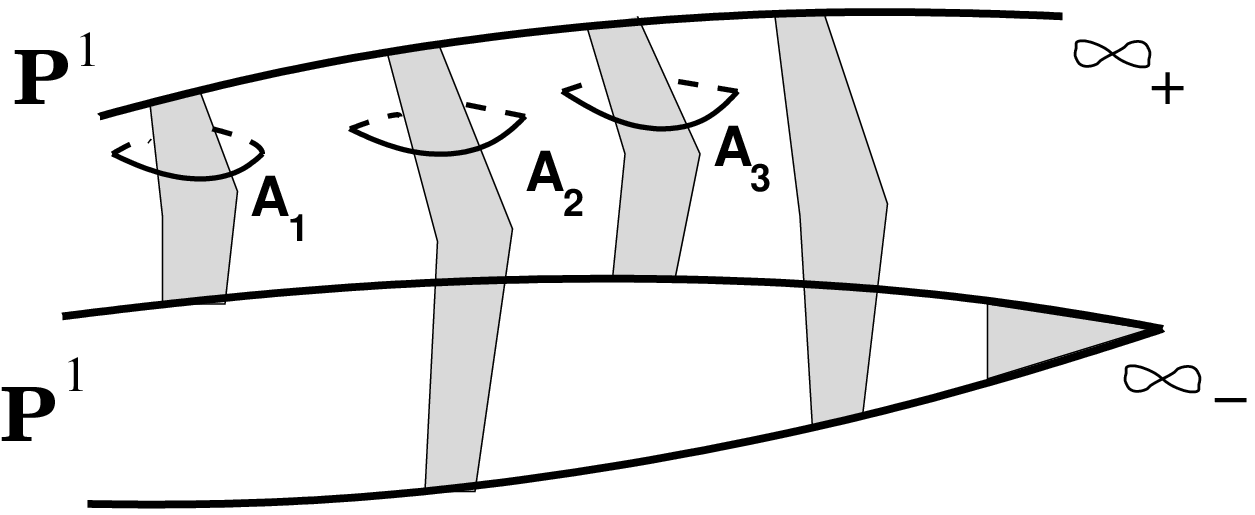}{70}{ Cubic curve as a cover of
$z$-plane. }

Let us now  present the curve \cuthree\ as a Riemann surface of
a multi-valued function ${\tilde z}(z)$. Then it can be thought as
a three-sheet cover of the complex $z$-plane. On the first, physical
sheet, there are no branch cuts at $z\to \infty$, as it follows from
the asymptotics \asyone. This asymptotics should be supplemented by
the "complex-conjugated" asymptotics
\eqn\asytwo{
z = {\overline W}'({\tilde z}) + O\left({1\over\tilde z}\right)}
on "unphysical" sheets. Then it is clear from \asyone\ and \asytwo\
that on the physical sheet at infinity ${\tilde z} \propto z^2$, while
on two unphysical sheets ${\tilde z} \propto \sqrt{z}$ and two
infinities on unphysical sheets are "glued" by a cut.

The branch points at $z$-plane are determined by zeroes of the
differential $dz$, or by $F_{\tilde z}=0$. Considering the simplest
non-degenerate case of the curve \cuthree\
\eqn\CUSIMP{z^2{\tilde z}^2 + az^3 + {\bar a}{\tilde z}^3
+  h = 0
}
it is easy to see that there are nine branch points in the $z$-plane
without infinity $z=\infty$
(of course, one comes to the same conclusion looking at the Cardano
formula, or from the index theorem, see below).

The structure of the curve can be then presented as on \cutre.  It is
clear from this picture that the curve can be presented as two copies
of ${\bf P}^1$ "glued" by four cuts, i.e. in general position it has
the genus $g=3$. There are two "infinities" $z=\infty$, ${\tilde
z}=\infty$, one of them is a branch point. We have shown schematically
the possible cuts and the corresponding choice of canonical ${\bf
A}$-cycles. In the classical situation \CUCLA\ one has two parabolas
intersecting at four points, and under quantum resolution these points turn
into four cuts connecting two ${\bf P}^1$'s in \cutre.

\ifig\cun{ 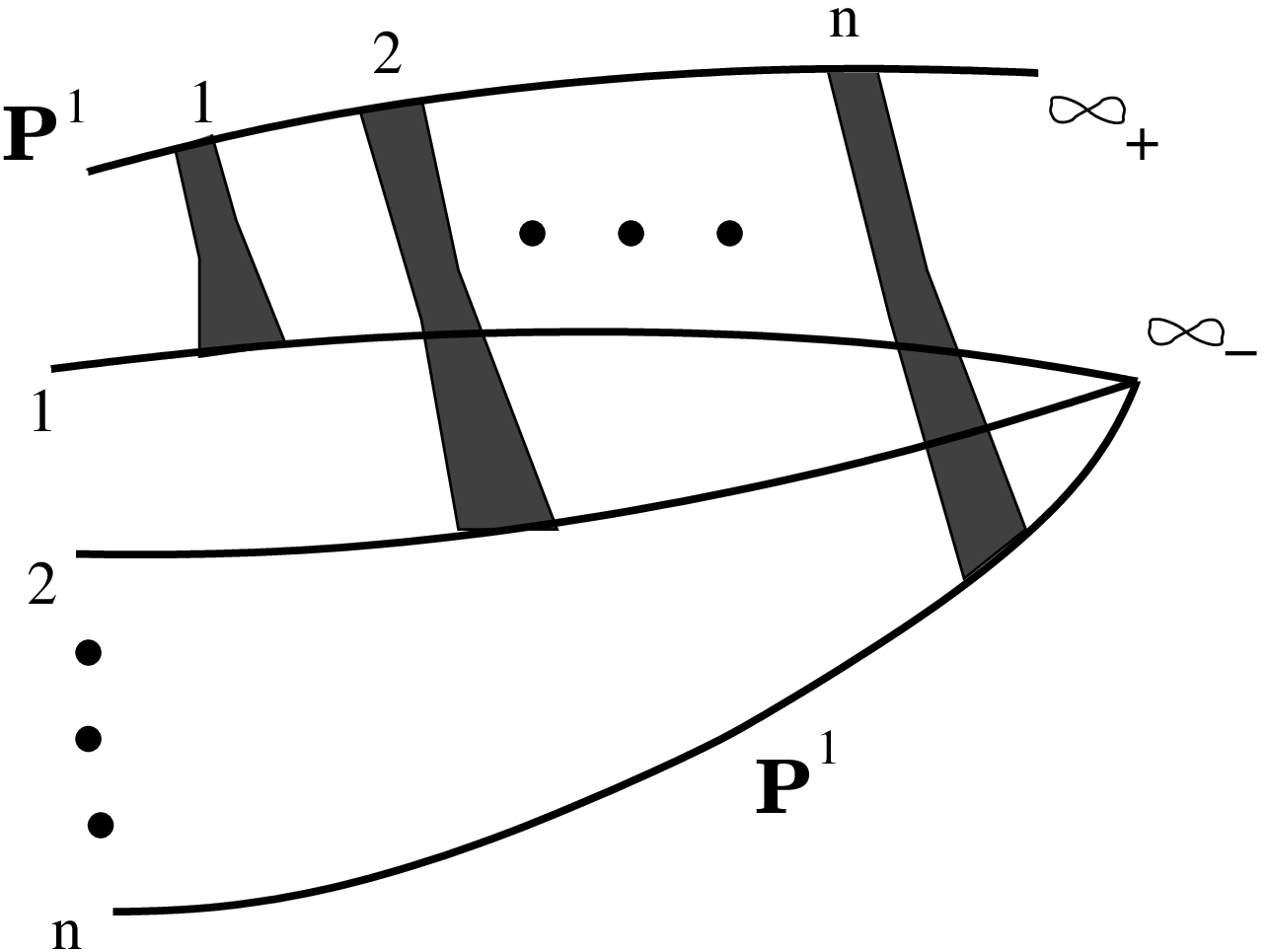}{60}{ Generic curve of the two matrix model as
a cover of the $z$-plane. In contrast to \cutre\ each fat line consists
of a stack of $n$ cuts.}

To conclude the picture, let us make few comments about the curve
\complcu\ for a potential  of a generic degree $n$, i.e. when
$W'(z)\sim z^n + \dots$. This curve (see \cun) can be again presented
as two ${\bf P}^1$'s glued by $n$ stacks of cuts. One of these ${\bf
P}^1$'s corresponds to the "physical sheet", the other one is glued at
the $\infty_-$ from $n$ copies of "unphysical" $z$-sheets. Each stack
consists of $n$ cuts, so their total number is $n^2$ among which one
can choose $n^2-1$ independent, whose number is equal to the genus of
this Riemann surface.

The differential $dz$ has always a pole of the second order at $\infty_+$ on
the upper, or "physical" sheet, and  a pole of the order $n+1$ at
$\infty_-$ since $\left. z\right|_{\infty_-} \propto {\tilde z}^n + \dots$. It
gives altogether $n+3$ poles and from the Riemann-Roch theorem one concludes
that the number of branching points, or zeroes of $dz$ is equal to
\eqn\ZERDZ{
\# (dz=0) = n+3 + 2(n^2-1)-2 = 2n^2 + n -1 }
reproducing nine for $n=2$. In general position this gives
exactly $2n^2$ branch points, producing simple cuts and $(n-1)$ ramification
points, connected by cuts with $\infty_-$.

\ifig\double{ 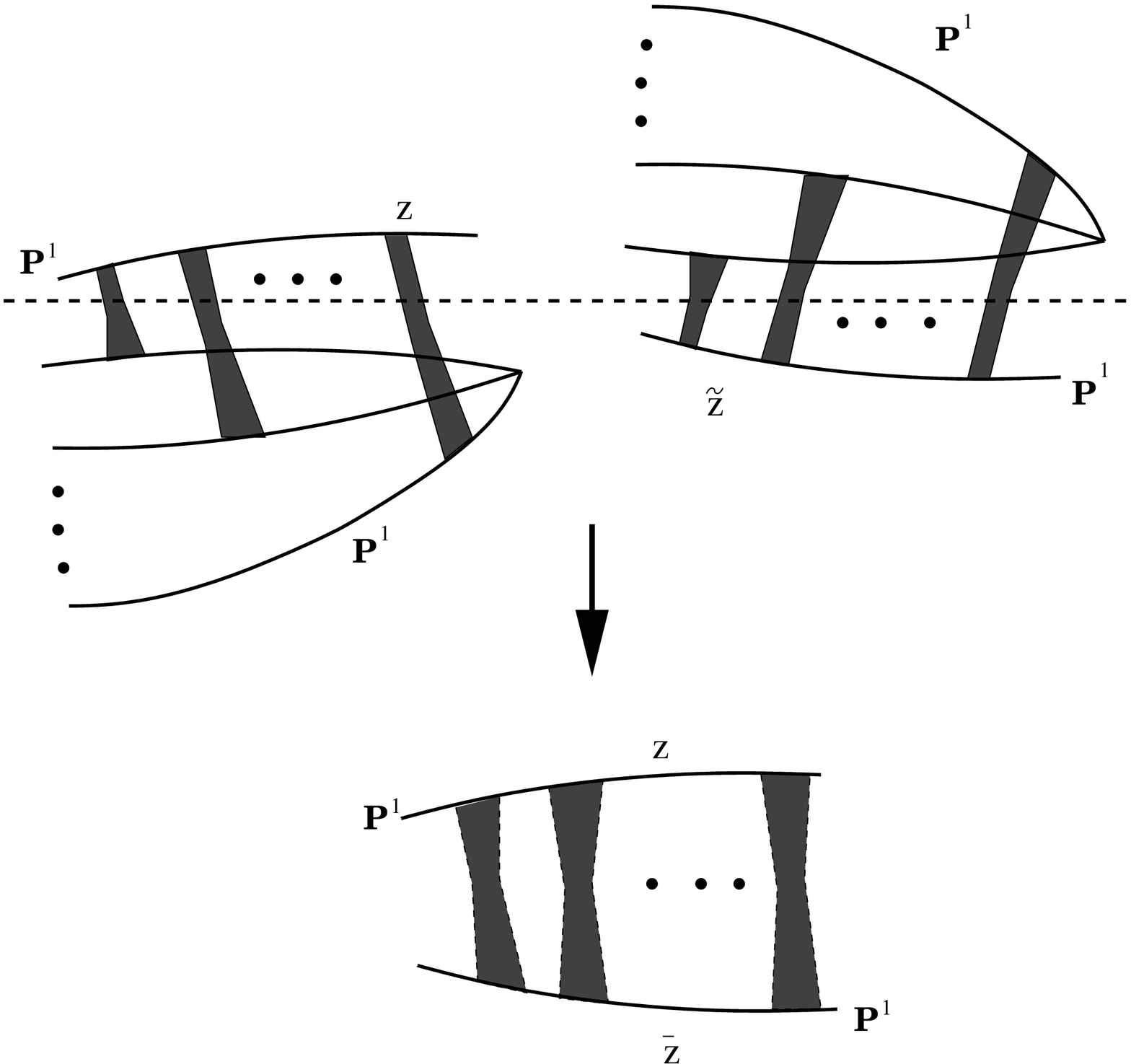}{90}{ Generic curve of the two matrix model as a
double of $z$ and ${\bar z}$ planes. One takes the Riemann surface of
the function ${\tilde z}(z)$, as on \cun, and its "mirror" Riemann
surface of the function $z(\tilde z)$ which possesses the same
structure. Cutting the physical sheets one may glue them together
along the (real) curves $\bar z = \tilde z(z)$.}

Finally, let us point out that the structure of the curve \complcu\ and
\cun\ is consistent with the structure of the "double" in $z$ and ${\bar z}$
variables, explicitly seen in the one-support solution \KKMWZ\ and
proposed to be the feature of the multi-support solutions by Krichever
\KriUP. Indeed, the eq. \complcu\ is "symmetric" with respect to $z$
and ${\tilde z}$ variables, so instead of the picture on \cun, one can
draw a "dual" picture of a $(n+1)$-sheet cover of the ${\tilde
z}$-plane. These dual pictures can be combined together as on
\double. Cutting "physical" sheets from both pictures one may glue
them together, forming a double with involution $z\leftrightarrow{\bar
z}$. The only delicate point is that these $z$ and ${\bar z}$ sheets
should be glued together along the boundaries of the spots where
${\tilde z}(z) = {\bar z}$ and vice versa, in contrast to the picture
of \cun, where the sheets are glued along the cuts on Riemann surface
of the multi valued function ${\tilde z}(z)$ defined by solution to
eq. \complcu. Both sheets of the double (the lower picture at \double)
generally have $n^2$ spots.

\ifig\cutspo{ 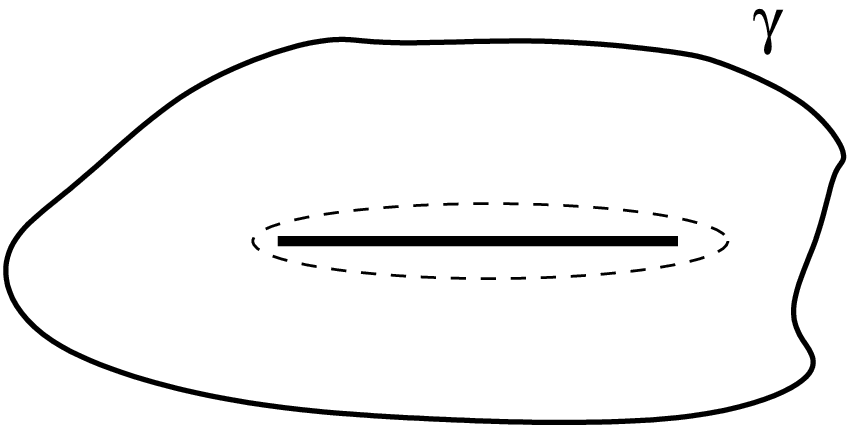}{50}{ The boundary of the spot $\gamma$ and a cut of
a multi-valued function $\tilde z(z)$ inside the spot. On $\gamma$ one has
an equality $\bar z = \tilde z(z)$ but this is not true on the cut.}

The difference between the boundary of a spot and a cut of a
multi-valued function is demonstrated on \cutspo. One obviously gets
the following relations for the two-dimensional and contour integrals
\eqn\RHODS{
\int_{\rm spot} dz\wedge d{\bar z} = \oint_\gamma \bar z dz =
\oint_\gamma \tilde z dz = \oint_{\rm cut} \tilde z dz }
which clarify the equivalence of two pictures on \double. The
relations \RHODS\ allow to endow the complex curve \complcu\
(or \NONSY\ in the asymmetric case) with a
meromorphic generating differential $\tilde z dz$.

\subsec{Degenerate curves}

Let us now discuss how the curve \complcu\ can be degenerated. The
(smooth) genus $g=n^2-1$ of the curve \complcu\ decreases if there exists
nontrivial solution to the system of equations
\eqn\DEGENR{
F(z,{\tilde z}) = 0, \ \ \ \ \ dF = 0}
It imposes certain constraints to the coefficients of $f_{ij}$ of the
equation \complcu, which can be found, say computing the resultant of
the equations \DEGENR. However, these constraints cannot be really
resolved in a general position.

To get an idea how the curve \complcu\ can be degenerated consider first
the cubic case \cuthree\ and let us put all coefficients of this equation
to be real. Then, it is easy to see that it can be rewritten in the form
\eqn\ELLW{
Y^2 + aX^3 + cX^2 + qX + h - {1\over 4}\left((3a-b)X + 2c-f\right)^2
\equiv Y^2+P(X) = 0.}
%
\ifig\cutrto{ 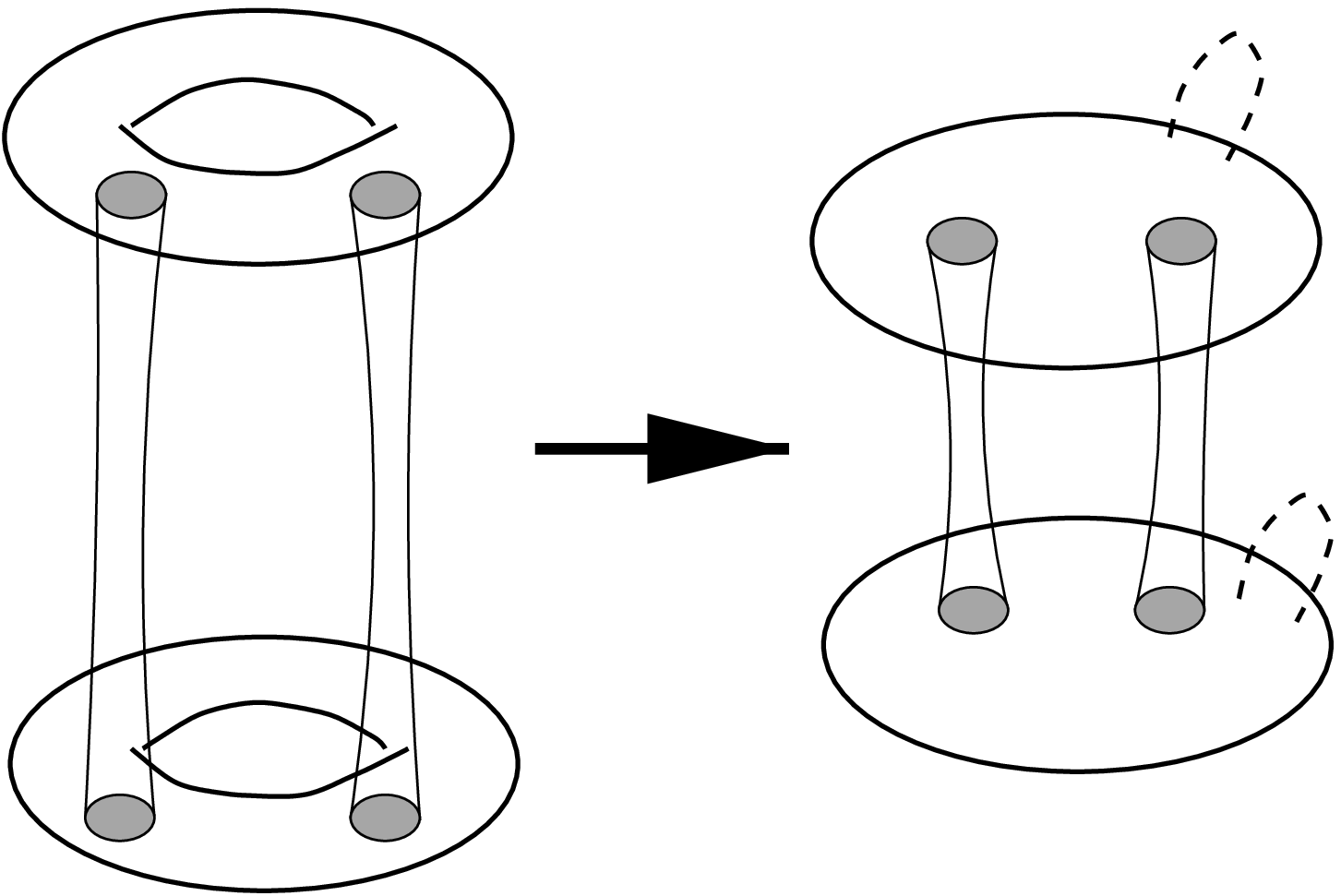}{60}{The curve \cuthree\ as double
cover of the torus. When the torus \ELLW\ degenerates, the genus
$g=n^2-1=3$ curve \cuthree\ degenerates into the curve of $g_{\rm
red}=n-1=1$.}
where
\eqn\COVER{
X = z + {\tilde z}, \ \ \ \ \
Y = z{\tilde z} - {1\over 2}\left((3a-b)X + 2c-f\right)}
One may "tune" for simplicity the coefficients of the potential
\coeff\ to get $3a=b$ and $2c=f$. The formulas \COVER\ show that our
curve \complcu\ can be presented as a double cover of the torus
\ELLW\ with four branch points which are solutions to the equation \cuthree\
under the substitution ${\tilde z}=z$, where the transformation
\COVER\ becomes singular. Hence, the curve \cuthree\ can be also presented
(in addition to \cutre) as two tori glued by two cuts (see \cutrto).

Now it becomes clear how this picture can be degenerated. Rewriting
equations \DEGENR\ as
\eqn\FDF{ \eqalign{
F_{\tilde z} = zF_{Y} + P'(X)
\cr
F_{\tilde z} = {\tilde z}F_{Y} + P'(X)
}}
one immediately finds that they lead either to $z={\tilde z}$ or to
$F_Y=0$ and $P'(X)=0$. In the second case the torus \ELLW\
degenerates, while $z={\tilde z}$ leads to degeneration of the cover
of this torus. We will be more interested in the degeneration of the
torus since, for example, it corresponds to filling of the "correct"
vacua (real eigenvalues) in the perturbative picture considered in
section 2.

When the torus degenerates into a rational curve, one gets the Riemann surface
\cuthree\ presented as a double cover of this rational curve with two cuts,
i.e. as a Riemann surface of genus $g=1$ with smooth handles of tori
degenerated into (a pair of) singular points (see \cutrto).

The equations of degeneration of the torus \ELLW\ can be easily
written using the conditions for the double root of the polynomial
$P(X)$.  Explicitly these conditions acquire the form of the
discriminant of $P(X)$ or the resultant of the two polynomials $P(X)$
and $P'(X)$.

Now, in the general case  \complcu\ with real coefficients
the substitution analogous to \COVER\ brings
it to the form
\eqn\QMM{
Y^n + X^{n+1} + \dots = 0 }
%
\ifig\newtxy{ 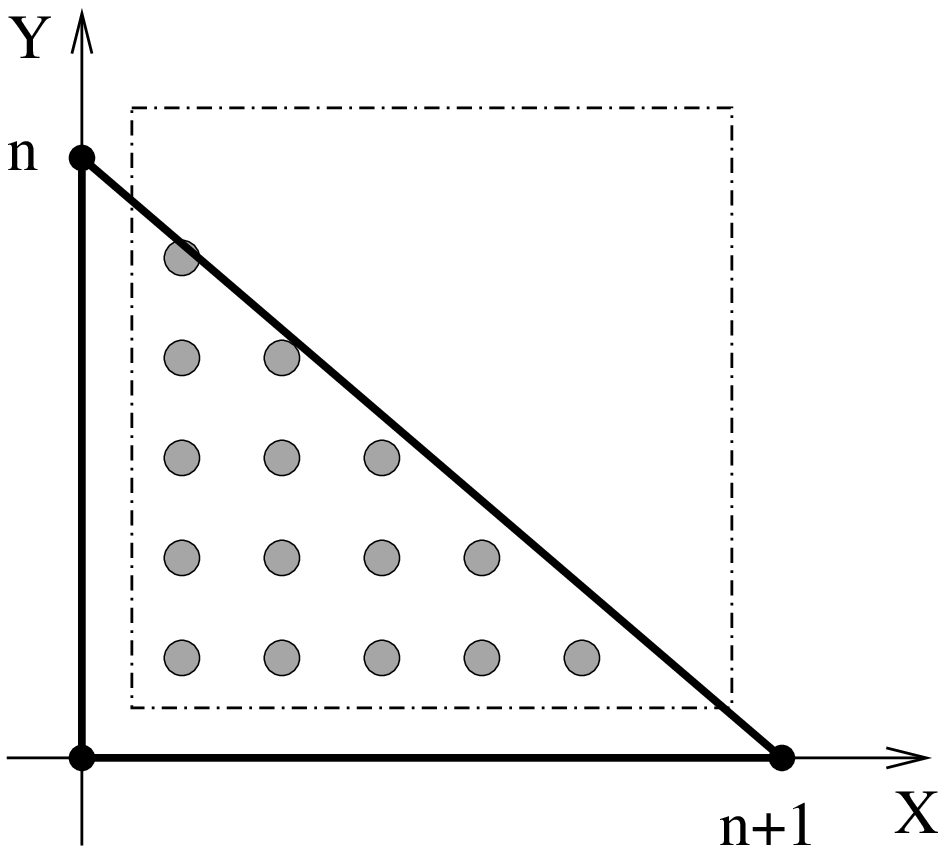}{60}{The Newton polygon for the curve \QMM\
gives the genus $g_\ast = {n(n-1)\over 2}$.}
where by dots we denoted monomials of lower degrees in $X$ and $Y$,
and there are no "mixed" terms in the eq. \QMM. The genus of the curve
\QMM\ can be again easily computed by the Newton polygon, which gives
\eqn\GQMM{
g_\ast = {n(n-1)\over 2} }
In the same way one may present the generic curve of the two matrix model
\complcu\ as a double cover of \QMM\ with $2n$ branch points.
Indeed, the Riemann-Hurwitz formula
\eqn\RH{
2-2g = \#\ S\cdot(2-2g_0) - \#\ B.P. }
where $\#\ S$ is number of sheets and $\#\ B.P.$ is number of branch
points, gives for $g=n^2-1$ and $g_0=g_\ast$ exactly $\#\ B.P.=2n$. It
means that the generic curve of the two matrix model \complcu\ can be
presented as a double cover of the curve \QMM\ with $n$ cuts, and when
the curve \QMM\ degenerates into a rational one, the curve \complcu\
has the genus
\eqn\GRED{
g_{\rm red} = n-1 }
%
\ifig\cucomm{ 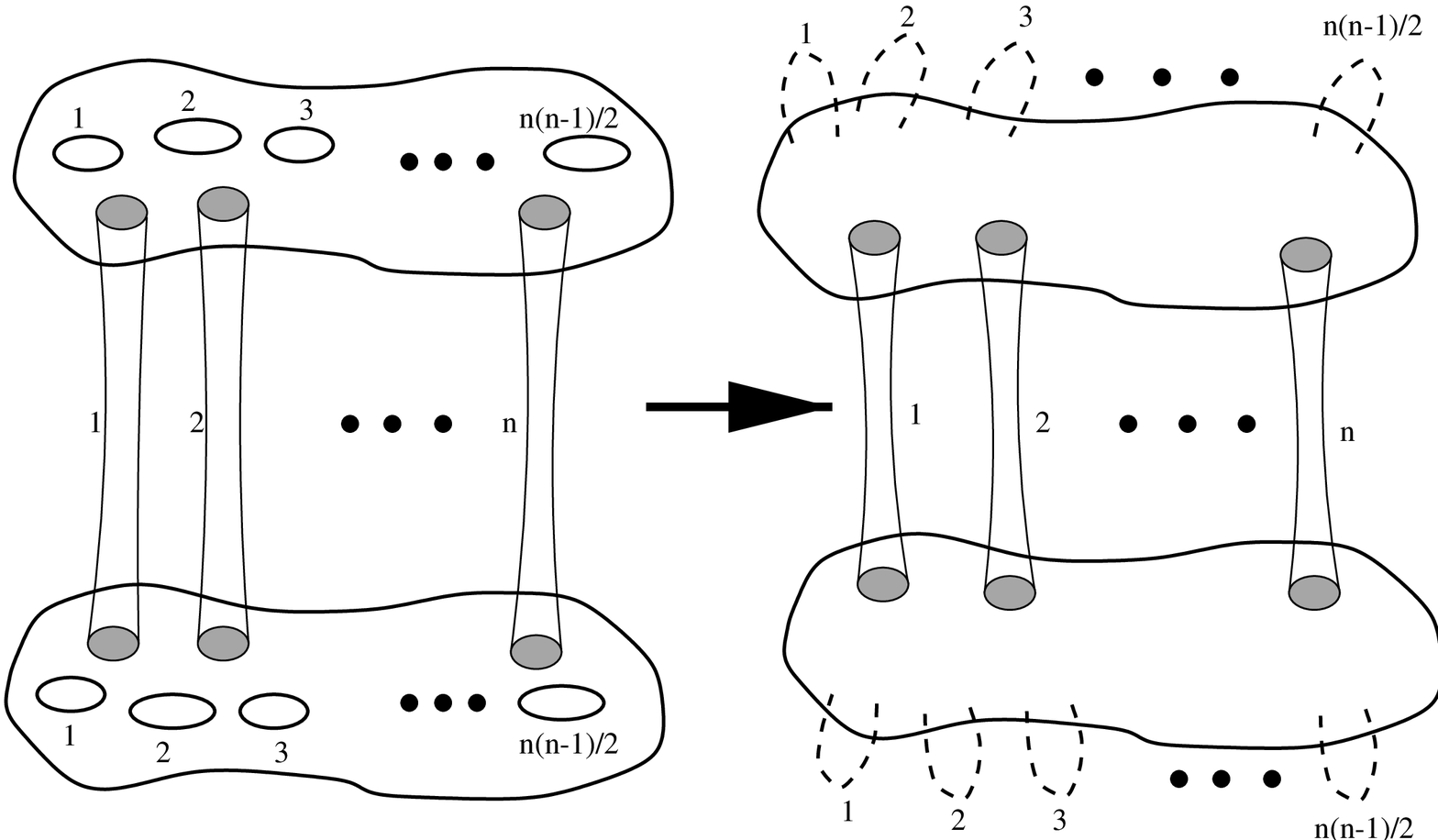}{100}{The general curve \complcu\ as a double
cover of the curve \QMM\ with a genus $g_\ast = {n(n-1)\over 2}$.
Similarly to \cutrto, when the curve \QMM\ completely degenerates
into a rational curve the curve of two matrix model
\complcu\ degenerates into the curve of genus $g_{\rm red}=n-1$.}

\subsec{Rational degenerations}

Let us finally say a few words about the  rational degenerations of
\complcu, i.e. when its (smooth) genus vanishes. A particular example of
such a totally degenerate curve is given by the "classical" curve
\CUCLA, but the rational case can be easily studied for the generic
values of coefficients in \complcu, i.e. without any reality
restriction.

In such situation eq. \complcu\ can be resolved via the (generalized)
conformal map
\eqn\COMAP{ \eqalign{
z = rw + \sum_{k=0}^n {u_k\over w^k}
\cr
\tilde z = {r\over w} + \sum_{k=0}^n \bar u_k w^k
}}
and the substitution of \COMAP\ into \complcu\ gives a system of
equations, expressing {\it all} coefficients $f_{ij}$ in terms of
parameters of the conformal map \COMAP.

Indeed, substituting \COMAP\ into \complcu\ and computing the residues
one finds that the expressions
\eqn\RESEQ{
R_l[F]={\rm res}\left({dw\over w}w^lF(z(w),\tilde z(w))\right) =0 }
for $l= -n(n+1),\dots,n(n+1)$ form a triangular system of equations
onto the coefficients $f_{ij}$. It means that each of the equations
\RESEQ\ is linear in one of the coefficients, and can be resolved step
by step, starting from the ends of the chain.

For the cubic potential ($n=2$) the solution is
$$
a =  - {   {r^{2}}\over{{  u_2}}}
$$
$$
b = {   {{  u_1}\,r}\over{{  u_2}}}  - 2\,{  {\bar u}_0}
$$
$$
c =  - {  {{  u_1}\,r\,{  {\bar u}_0}}\over{{  u_2}}}
 + {  {\bar u}_0}^{2} - 2\,r\,{  {\bar u}_1} + 3\,{  {r^{2}
\,{  u_0}}\over{{  u_2}}}  - {  {r^{3}\,{  {\bar u}_1}}\over {
{  {\bar u}_2}\,{  u_2}}}
$$
$$
f = r^{2} - 2\,{  u_2}\,{  {\bar u}_2} + 4\,{  u_0}\,{  {\bar u}_0} +
{  {r^{4}}\over{{  {\bar u}_2}\,{  u_2}}}  - {  u_1}\,
{  {\bar u}_1} - 2\,{  {r\,{  {\bar u}_1}\,{  {\bar u}_0}}\over{{
{\bar u}_2}}}  + {  {r^{2}\,{  {\bar u}_1}\,{  u_1}}\over{{  {\bar u}_2
}\,{  u_2}}}  - 2\,{  {r\,{  u_0}\,{  u_1}}\over{
{  u_2}}}
$$
$$
q =  - 3\, {{r^2\,  u_0^2}\over{  u_2}}  + 2u_2{\bar u}_2{\bar u}_0 -
u_2 {\bar u}_1^2 - 2\,{  {r^{2}\,  {\bar u}_1^2}\over{{  {\bar u}_2}}}
 - {  {r^{4}\,{  {\bar u}_0}}\over{{  {\bar u}_2}\,{  u_2}}}
 - 3\,r\,{  {\bar u}_2}\,{  u_1} + {  {\bar u}_0}\,{  u_1}\,{  {\bar u}_1}
$$
$$
+ 4\,
{  u_0}\,r\,{  {\bar u}_1} - 2\,{  {\bar u}_0}^{2}\,{  u_0} - {  {
\,r\,{  {\bar u}_1}}{  u_1}^{2}\over{{  u_2}}}  - r^{2}\,{  {\bar u}_0} + 2\,
{  {r\,\,{  u_1}{  {\bar u}_0}\,{  u_0}}\over{{  u_2}}
}  + 3\,{  {r^{3}\,{  u_1}}\over{{  u_2}}}
$$
$$
 +
{  {r\,{  {\bar u}_1}\,{  {\bar u}_0}^{2}}\over{{  {\bar u}_2}}}  -
{  {r^{2}\,{  {\bar u}_0}\,{  u_1}\,{  {\bar u}_1}}\over{{
{\bar u}_2}\,{  u_2}}} + 2\,{  {{  u_0}\,r^{3}\,{  {\bar u}_1}
}\over{{  {\bar u}_2}\,{  u_2}}}
$$
$$
h =  - {  {r^{6}}\over{{  {\bar u}_2}\,{  u_2}}
}  + {  {r^{2}\,{  {\bar u}_0}^{3}}\over{{  {\bar u}_2}}}  +
{  {\bar u}_2}\,{  {\bar u}_0}\,{  u_1}^{2} - {  {{  u_0}
\,r\,{  {\bar u}_1}\,{  {\bar u}_0}^{2}}\over{{  {\bar u}_2}}}  +
{  u_2}\,{  u_0}\,
{  {\bar u}_1}^{2} - 3\,{  {r^{3}\,{  {\bar u}_0}\,{  {\bar u}_1}
}\over{{  {\bar u}_2}}}  - {  {{  {\bar u}_2}\,{  u_1}^{3}\,r}\over{
{  u_2}}}
$$
$$
 - {  {{  u_2}\,r\,{  {\bar u}_1}^{3}}\over{
{  {\bar u}_2}}}  + {  {r^{2}\,{  u_0}^{3}}\over{{  u_2}
}}  + 2\,{  {r^{4}\,{  {\bar u}_1}\,{  u_1}}\over{{  {\bar u}_2
}\,{  u_2}}}  + {  {{  u_0}\,{  {\bar u}_0}\,{
{\bar u}_1}\,r^{2}\,{  u_1}}\over{{  {\bar u}_2}\,{  u_2}}}  - {
{r^{3}\,{  u_1}\,{  {\bar u}_0}^{2}}\over{{  {\bar u}_2}\,{  u_2}}}  -
{  {r^{3}\,{  {\bar u}_1}\,{  u_0}^{2}}\over{{  {\bar u}_2}\,
{  u_2}}}  - {  {{  u_1}^{2}\,{  {\bar u}_1}^{2}\,r
^{2}}\over{{  {\bar u}_2}\,{  u_2}}}
$$
$$
 + 3\,r^{4} + {  {\bar u}_0}^{2}\,{  u_0}^{2} +
{  {{  u_1}^{2}\,{  {\bar u}_1}\,{  u_0}\,r}\over{{
u_2}}}  - 2\,{  {\bar u}_0}^{2}\,{  u_1}\,r + 2\,{  {
{  u_1}^{2}\,r^{2}\,{  {\bar u}_0}}\over{{  u_2}}}  + {
{{  u_1}\,{  {\bar u}_1}^{2}\,r\,{  {\bar u}_0}}\over{{  {\bar u}_2}}}
+ r^{2}\,{  {\bar u}_0u_0}
$$
$$
 - {  u_1}\,{  {\bar u}_1}\,{  u_0}\,{  {\bar u}_0} +
{  {{  u_0}\,{  {\bar u}_0}\,r^{4}}\over{{  {\bar u}_2}\,{
u_2}}}  + {  u_2}^{2}\,{  {\bar u}_2}^{2} - {  {{
u_1}\,r\,{  {\bar u}_0}\,{  u_0}^{2}}\over{{  u_2}}}
- 2\,r\,{  {\bar u}_1}\,
{  u_0}^{2} + 2\,{  {{  u_0}\,r^{2}\,{  {\bar u}_1}
^{2}}\over{{  {\bar u}_2}}}  - 3\,{  {r^{3}\,{  u_0}\,
{  u_1}}\over{{  u_2}}}
$$
$$
 - {  {\bar u}_1}\,{  u_1}\,{  {\bar u}_2}\,{  u_2} - 3\,r^{2}
\,{  {\bar u}_2}\,{  u_2} + 3\,{  u_2}\,r\,{  {\bar u}_1}\,{  {\bar u}_0} - 2\,
{  {\bar u}_2}\,{  u_2}\,{  {\bar u}_0}\,{  u_0} - {  {\bar u}_1}\,r^{2}\,{  u_1
} + 3\,{  u_0}\,r\,{  {\bar u}_2}\,{  u_1}
$$
together with the "complex conjugated" expressions for ${\bar a}$,
${\bar b}$, ${\bar c}$ and ${\bar q}$, where one should replace $u_k$
by $\bar u_k$ and vice versa. Resolving \RESEQ\ one gets the explicit
description of the rational degeneration of the curve \complcu\ in
terms of the coefficients of conformal map \COMAP. However, in general
situation they are only implicitly defined through the parameters of
the potential $V(z,{\bar z})$.

\newsec{Quasiclassical tau-function}

Let us now define the partition function for the two matrix model
\variF\ in terms of the quasiclassical tau-function introduced in
\KriW. First, we discuss a simpler example of the one matrix model
 and then turn to the particular features of the two-matrix case.  The
hyperelliptic curve of the one matrix model was first discussed in
\David, and recently, in  the most general form including all the
extrema of the potential, in \DV\ (see also \GDKV\ and \FERRA).

\subsec{One-matrix model}

The complex curve of the one matrix model
\eqn\OMM{ {\cal Z}=\int d\Phi e^{ \Tr W_n(\Phi)}  }
with the  model potential
\eqn\mmpot{
W_n'(\Phi) = \sum_{k=1}^n kt_k \Phi^{k-1} }
%
comes from the very simple loop equation $G^2 + 2W_n'(\l)G - f(\l)=0$
(see, for example, \Migdal\ or \KAZMP). It is always hyperelliptic,
i.e. can be rewritten in the form
\eqn\dvc{
y^2 = W'_n(\lambda)^2 + f(\lambda)}
with $y=G+W'_n$ and the moduli hidden in the coefficients of
the polynomial
\eqn\f{
f(\lambda) = \sum_{k=0}^{n-1}f_k\lambda^k. }
The generating differential is chosen as
\eqn\dvds{
dS^{1MM} = {1\over 2\pi i}\ yd\lambda }
and additional variables, corresponding to the eigenvalue filling
numbers, can be introduced through its periods
\eqn\DVper{
S_i = \oint_{A_i}dS^{1MM} }
directly related to the integrals of density over the eigenvalue
supports.  Then
\eqn\canoDV{ \eqalign{
{\p dS^{1MM}\over\p S_i} = d\omega_i
\cr
\oint_{A_i}d\omega_j = \delta_{ij}, }}
where the derivatives are taken at fixed coefficients $\{ t_l\}$ of
the potential $W_n'(\lambda)$ \mmpot. As usual, the periods dual to
\DVper\ are given by the integrals over dual cycles
\eqn\DVFz{
\Pi^{1MM}_i = \oint_{B_i}dS^{1MM}. }
To complete the set of parameters of the model, we have to add to the
filling numbers \DVper\ and coefficients of the potential
\mmpot\ the variable
\eqn\tz{
{\rm res}_{\infty_+}\left(dS^{1MM}\right) = - {\rm
res}_{\infty_-}\left(dS^{1MM}\right) = {f_{n-1}\over 2t_n} \equiv t_0,
}
so that
\eqn\bipole{
{\p dS^{1MM}\over \p t_0} =t_n {\lambda^{n-1} d\lambda\over y} +
\hf \sum_{k=0}^{n-2}{\p f_k\over\p t_0}{\lambda^k d\lambda\over y}. }
The dependence of $\{ f_k\}$ with $k=0,1,\dots,n-2$ upon $t_0$ is
fixed by
\eqn\van{
\oint_{A_i}\left(t_n{\lambda^{n-1} d\lambda\over y} +
\hf\sum_{k=0}^{n-2}{\p f_k\over\p t_0}{\lambda^k d\lambda\over y}\right)=0 }
which gives for $i=1,\dots,n-1$ exactly $n-1$ relations on
$f_0,f_1,\dots,f_{n-2}$.  The bipole differential \bipole\ can be also
rewritten as
\eqn\bp{
d\Omega_{\pm} = {\p dS^{1MM}\over \p t_0} = d\log
\left({E(P,\infty_+)\over E(P,\infty_-)}\right) }
where $E(P,P')$ is the Prime form on \dvc. Obviously, the differentials
\bp\  obey the properties
\eqn\obv{ \eqalign{
{\rm res}_{\infty_+}d\Omega_{\pm} =
- {\rm res}_{\infty_+}d\Omega_{\pm} = 1
\cr
\oint_{A_i}d\Omega_{\pm} = 0,\ \ \ \ \ \ i=1,\dots,n-1 }}
To complete the setup
one should also add to \DVFz\ the following formula
\eqn\dfdt{
\Pi_0 = \int_{\infty_-}^{\infty_+}dS^{1MM} }
which can be regularized in the usual way presenting the puncture at
infinity as a degenerate handle.  The partition function of the
multi-cut solution of the 1MM is defined now in terms of the
quasiclassical tau-function $\CF^{1MM}$ obeying the equations
\David,\DV
\eqn\FOMM{ \eqalign{
{\p{\cal F}^{1MM}\over \p S_i^{1MM}} = \Pi_i^{1MM}
\cr
{\p{\cal F}^{1MM}\over \p t_0} = \Pi_0
}}
In the papers \DV, instead of $t_0$ the parameter $\tilde S = t_0 -
\sum_{i=1}^{n-1}S_i \equiv S_n$ was used. This is a non-standard
definition of homology basis on \dvc\ and it gives rise to the
divergences at infinities.  However, the basis of \DV\ is related to
the canonical one by a linear change of variables, where no
divergences appear (except of the trivial one in \dfdt) and the
integrability of \FOMM\ (the symmetry of second derivatives) follow
from the Riemann bilinear relations, including the symmetry of period
matrix of \dvc.

\subsec{Two-matrix model (general
 potential)}

In the same way, the filling numbers can be defined for the
two matrix model
\eqn\pertmm{
S_i  = {1\over 2\pi i}\int_{i-\rm th\ spot}dz\wedge d{\bar z} =
{1\over 2\pi i}\oint_{A_i} {\tilde z}dz = \oint_{A_i}dS^{2MM},}
i.e. as periods of the generating differential
\eqn\diffmm{
dS^{2MM} = {1\over 2\pi i}\ {\tilde z}dz}
under the appropriate choice $\{ A_i\}$ for the basis of $A$-cycles on
the Riemann surface \complcu, (or \NONSY). This is illustrated by
\cun\ (or by its
particular cubic case \cutre), taking into account \cutspo\ and
eqn. \RHODS.  From \pertmm\ one still gets in the same way the analogs
of the formulas \canoDV
\eqn\CANOTMM{ \eqalign{
{\p dS^{2MM}\over\p S_i} = d\omega_i
\cr
\oint_{A_i}d\omega_j = \delta_{ij} }}
where the canonical holomorphic differentials (now on the curve
\complcu) are certain linear combinations (with moduli dependent
coefficients) of $g=n^2-1$ "lower degree" holomorphic differentials
\vij.

The derivatives of \diffmm\ with respect to the coefficients of the equation
\complcu\ can be computed in the  standard way. Choosing $z$ as a
covariantly constant function one writes for \complcu
\eqn\deltaF{
F_{\tilde z}\delta{\tilde z} + \delta F = 0  }
where $\delta F \equiv \sum \delta f_{ij}z^i{\tilde z}^j$ is a variation
of only the coefficients of \complcu. Then the  variation of
\diffmm\ gives rise to
\eqn\vardS{
\delta{\tilde z} dz = - \delta F {dz\over F_{\tilde z}} =
- \sum \delta f_{ij} z^i{\tilde z}^j {dz\over F_{\tilde z}}     }
Expression \vardS\ contains a decomposition of the variation of the
meromorphic differential \diffmm\ over some basis of meromorphic and
holomorphic differentials on the curve \complcu. It is easy to check
that the coefficients $f_{ij}$ corresponding to the meromorphic
Abelian differentials of the second kind can be expressed through the
parameters of the potential $V(z,{\bar z})$ of the two-matrix model,
namely, through the coefficients of its harmonic part $W(z) + {\bar
W}({\bar z})$. The corresponding relations follow from the fact that
the complex curve \complcu\ should satisfy the asymptotic expansion of
the branch
\eqn\branch{
{\tilde z} = W'(z) + O\left({1\over z}\right) =
\sum_{k=1}^{n+1}kt_kz^{k-1} + O\left({1\over z}\right)   }
which gives rise, e.g. to ${\bar a}^{-1} = - (n+1) t_{n+1}$ etc.

The rest of  coefficients $f_{ij}$ consists of the coefficient
corresponding to the third kind Abelian differential (with the
first-order pole) and the holomorphic differentials \vij. Fixing
the coefficients expressed through the parameters of the potential
in \vardS\ and taking appropriate linear combinations, one arrives
at \CANOTMM.

As it will be shown below, the dependence of the free energy \variF\
upon the filling numbers
\pertmm\ is defined by
\eqn\dptmm{
{\p {\cal F}\over\p S_i} = \oint_{B_i} dS^{2MM} = {1\over 2\pi
i}\oint_{B_i} {\tilde z}dz }
where $\{ B_i \}$ are the canonical dual cycles $A_i\circ
B_j=\delta_{ij}$.  The integrability of \dptmm\ follows from the
symmetry of the period matrix of the curve \complcu.

Now, as in the one matrix case, one should also introduce
\eqn\tztmm{
{\rm res}_{\infty_+}\left(dS^{2MM}\right) =
- {\rm res}_{\infty_-}\left(dS^{2MM}\right) =
 t_0 }
adding the bipole differential
\eqn\bptmm{
d\Omega_{\pm} = {\p dS^{2MM}\over \p t_0} = d\log
\left({E(P,\infty_+)\over E(P,\infty_-)}\right) =
z^{n-1}{\tilde z}^{n-1}{dz\over F_{\tilde z}} - \sum_{\rm holomorphic}
{\p f_{ij}\over\p t_0}dv_{ij} }
where coefficients ${\p f_{ij}\over\p t_0}$ are fixed, as in \van, by
$\oint_{A_i}d\Omega_{\pm} =0$. The variables \pertmm\ should be
directly identified with those introduced in \variF\ for
$\alpha=1,\dots,n^2-1$ while $S_{n^2} \equiv t_0 - \sum_i^{n^2-1}
S_i$. The second kind Abelian differentials are defined as derivatives
$$
d\Omega_k = {\p dS^{2MM}\over\p t_k}, \ \ \ \ \ \
d\bar\Omega_k = {\p dS^{2MM}\over\p \bar t_k}
$$
over the parameters of the matrix model potentials, and the dependence of
the coefficients $f_{ij}$ upon the coefficients of potential is fixed by
$\oint_{A_i}d\Omega_k = \oint_{A_i}d\bar\Omega_k = 0$.

The formulas \pertmm, \dptmm\ and \tztmm, together with the
(regularized) equation
\eqn\dfttz{
{\p {\cal F}\over\p t_0} = \int_{\infty_-}^{\infty_+} dS^{2MM} =
{1\over 2\pi i}\int_{\infty_-}^{\infty_+} {\tilde z}dz }
define the quasiclassical tau-function \KriW.

It is clear that this definition coincides in fact with the definition of
free energy of NMM or H2MM \variF. Indeed, using the formula \variF\
one can easily check that
\eqn\DPFUt{
{\p{\cal F}\over\p t_k} = \int z^k \rho(z,{\bar z})d^2z =
{1\over 2\pi i}\ {\rm res} \left( z^k {\tilde z} dz\right)
}
and
\eqn\DPFUS{
{\p{\cal F}\over\p S_\alpha} = v_\alpha
}
The quantities \DPFUS\ are in fact nothing but  linear combinations
of \dptmm\ and \dfttz, like in the case of the 1MM. It can be derived
by carefully treating the logarithmic integral $\int dz\wedge d{\bar
z}\log|z-z_\alpha|$, with $z_\alpha$ belonging to one of the
eigenvalue supports. Indeed from \vareq\ one can get for $v_\alpha -
V(z_\alpha,{\bar z}_\alpha)$
\eqn\LOGINT{ \eqalign{
\int dz\wedge d{\bar
z}\log|z-z_\alpha|^2 =
\sum_{\rm spots}\int dz\wedge d{\bar z}\log (z-z_\alpha) + c.c.
=
\cr
= \sum_{\rm boundaries}\oint \log (z-z_\alpha) {\bar z}dz + c.c. =
\sum_{\rm boundaries}\oint \log (z-z_\alpha) {\tilde z}dz + c.c. }}
%
\ifig\intlog{ 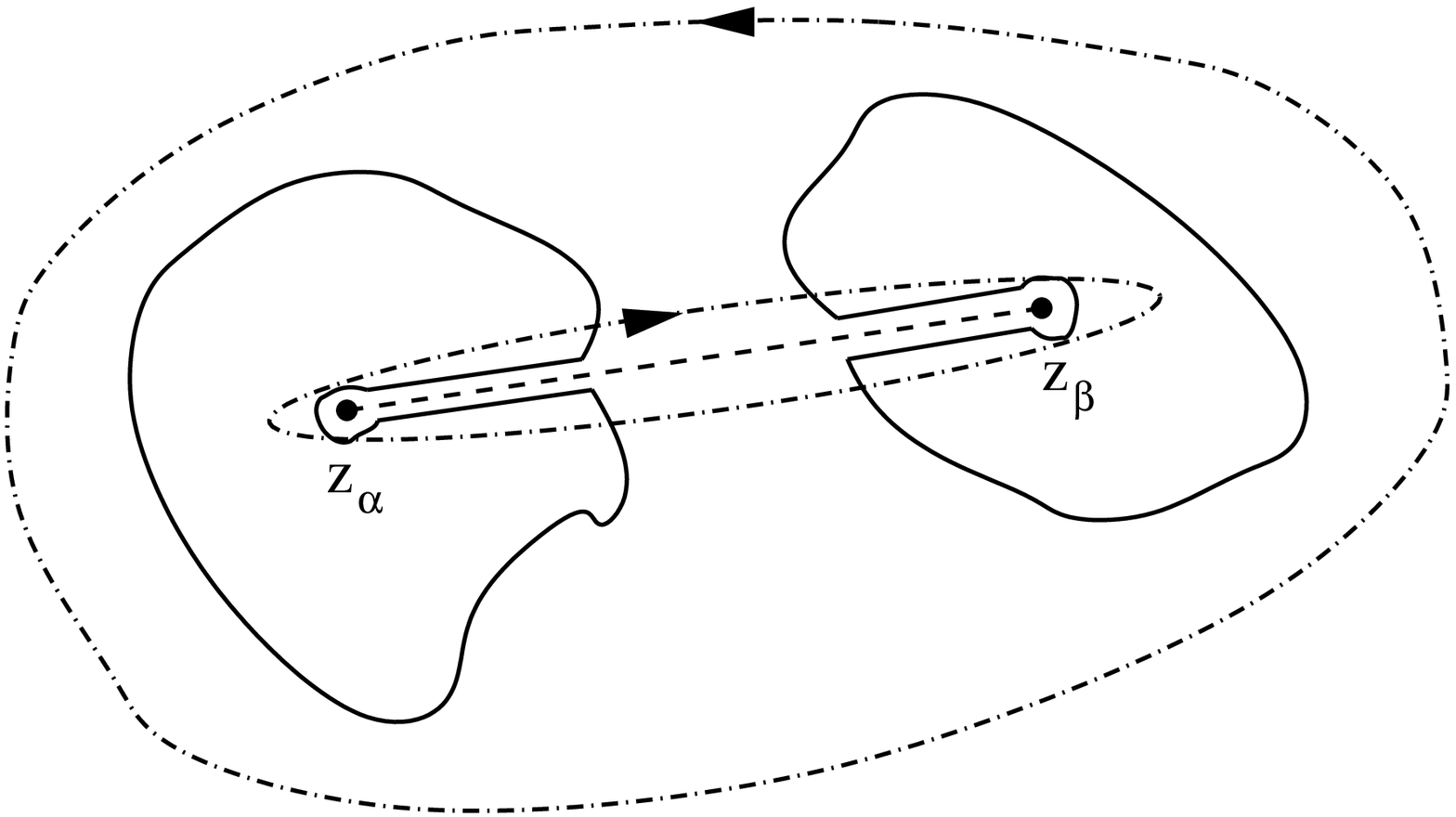}{70}{The integral in \LOGINT\ can be
transformed to the (linear combinations of) the integrals over the
cuts of logarithms, which turn into the $B$-periods of
$dS^{2MM}={1\over 2\pi i}{\tilde z}dz$.}
The integral in the r.h.s. of \LOGINT\ is an integral of the
multi-valued differential defined on a Riemann surface with two sorts
of cuts: the cuts of the function ${\tilde z}(z)$ and the cuts of the
logarithms. The integral over the spots can be transformed into
boundary integrals, where the boundaries should now include the
integrals along the branches of the logarithmic cuts (see \intlog),
the corresponding contributions to the two-dimensional integrals over
spots vanish. The latter ones are combined into the integrals around
all spots which can be deformed to the infinity, modulo the integrals
along the logarithmic cuts. Altogether these contributions gives rise
to the canonical $B$-periods of $dS^{2MM}={1\over 2\pi i}\tilde zdz$.
Finally, the situation appears to be quite similar to 1MM since
applying the procedure of "transfer" of an eigenvalue from one support
to another \DV\ one should leave intact the boundary of the spot
(since adding an eigenvalue directly to the boundary not only changes
the filling numbers, but also the shape of the domain, related to the
parameters of the potential \MaWZ). Instead, as in one matrix case
(where one may neglect this problem due to vanishing of the density of
eigenvalues at the "boundaries" of a cut) one should put the points
$z_\alpha$ to the branching points of the complex curve where
$dz=0$. After that, using \LOGINT\ and
$$
\left.{\p V(z,{\bar z})\over\p z}\right|_{dz=0} \propto \oint_{cut}
{\tilde z(z')dz'\over z-z'}
$$
one gets the formula \dptmm.


Indeed, let us calculate, for example, the difference
${\p {\cal F}\over\p S_i}-{\p {\cal F}\over\p S_j}$ by  procedure of the
``eigenvalue transfer'' from an endpoint $z_\beta=z''$ of the $\beta$-th
cut to an
endpoint $z_\alpha=z'$ of the  $\alpha$-th cut in $z$-plane.
It will be given by  difference of
the corresponding eigenvalue effective actions (see \NMMZ, \HMMEV)
at the saddle point:
\eqn\DIFFS{ \eqalign{ {\p {\cal F}\over\p S_i}-{\p {\cal F}\over\p S_j}=
-z'\tilde z'+z''{\tilde z}'' + W(z')- W(z'')+ \tilde W({\tilde z}')-
\tilde W({\tilde z}'')+ \cr +
\mathop{{{\sum}'}}\limits_{m=1}^{N}
\log{(z'-z_m)({\tilde z}'-{\tilde z}_m)\over
(z''-z_m)({\tilde z}''-{\tilde z}_m)}, }}
where ${\tilde z}'={\tilde z}(z')$, ${\tilde z}''={\tilde z}(z'')$, with the function ${\tilde z}(z)$ defined by the
algebraic curve \complcu\ (or \NONSY) of the two matrix model.

Passing to the continuum limit and introducing the resolvents $G(z)$ and
$\tilde G({\tilde z})$, as in \RES, we
rewrite the last term in \DIFFS\ as follows
$$
\oint_C {dz\over 2\pi i}\ G(z)\ \log\({(z'-z)\over (z''-z)}\)+
\oint_{\tilde C} {d{\tilde z}\over 2\pi i}\ \tilde G({\tilde z})\ \log\({({\tilde z}'-{\tilde z})\over ({\tilde z}''-{\tilde z})}\)
$$
where the contours $C$ and $\tilde C$ encircle all the eigenvalue
supports -- the cuts in the $z$ and ${\tilde z}$ planes, respectively (these
cuts or their stacks are depicted in
\double). It is important that,
according to the definition of the sums in the last term of \DIFFS, the
contours do not encircle the logarithmic cuts  along $(z',z'')$ and
$({\tilde z}',{\tilde z}'')$ intervals.

Blowing up the contours $C,\tilde C$ we will encircle only the logarithmic
cuts (note that there are no poles at infinity). Calculating
discontinuity along these cuts we reduce the contour integrals to the
ordinary ones:
\eqn\DIFFN{  {\p {\cal F}\over\p S_i}-{\p {\cal F}\over\p S_j}=
-z'{\tilde z}'+z''{\tilde z}'' +
\int_{z''}^{z'} dz\ {\tilde z}(z) + \int_{{\tilde z}''}^{{\tilde z}'}
d{\tilde z}\  z({\tilde z}), }
the potentials being absorbed into the  functions ${\tilde z}(z)
=W'(z)+G(z)$ and
$z({\tilde z})=\tilde W'({\tilde z})+\tilde G({\tilde z})$, according to
the saddle point equations
of the two matrix model.

Integrating by parts in the last term of
\DIFFN\  (note that  the last integral after the change of variables
goes along  the unphysical sheets of the curve \cun)
we finally get the integral over the dual $B_{ij}$ cycle
\eqn\DIFIN{  {\p {\cal F}\over\p S_i}-{\p {\cal F}\over\p S_j}=
 \oint_{B_{ij}} {{\tilde z}dz\over 2\pi i}  }
which is equivalent to the equation \dptmm\ defining the geometry of the
planar limit in two matrix model.

\subsec{Explicit form of the two-support solution for the real
cubic potential}

In the rest of this section we will study in detail the degenerate
case of a cubic potential with real coefficients, having only two
eigenvalue supports (on the real axis), instead of four.  The
degeneration greatly simplifies the calculations. The period integrals
can be even rewritten in terms of elliptic integrals.

Indeed, the degenerate torus \ELLW\ can be presented as
\eqn\PDG{
Y^2+P(X) =Y^2 -{1\over g}\left(X-X_1\right)^2\left(X-X_2\right)=0 }
and the new parameters $X_1$ and $X_2$ are defined by the
discriminant equations
\eqn\PDER{\eqalign{
P(X)&= -{1\over g}X^3+\left({T\over g}-{9\over 4g^2}\right)X^2
+\left(q+{3T\over g^2}-{3f\over g}\right)X+h-{1\over 4} \left({2T\over
g}-f\right)^2 =0\cr
P'(X) &= -{3\over g}X^2+\left({2T\over g}-{9\over
2g^2}\right)X +\left(q+{3T\over g^2}-{3f\over g}\right) = 0 }}
where we used \ELLW\ and \coeff.  This system of eqs. fixes, for example,
$h$ which is now not independent and can be expressed through $q$ and $f$.

Note that we have the expansions
\eqn\XEXP{\eqalign{
X_1 = - {1\over g} + \delta X_1 = - {1\over g} + O(\delta f,\delta q)
\cr
X_2 = - {1\over 4g} +T + \delta X_2 = - {1\over 4g} + T + O(\delta f,\delta q),
}}
where
\eqn\DEVI{ \eqalign{
\delta f &= f - {1\over g^2}
\cr
\delta q &= q + {T\over g^2}
}}
are deviations of moduli from their classical values in
eq. \CUCLA. The ``classical'' value of $X_1$ in \XEXP\ precisely
corresponds to the classical solution of \CUCLA, of course different
from diagonal the $z=\tilde z$.

Substituting $z=\tilde z$ into eq. \cuthree\ one gets the fourth-order
equation
\eqn\FOUROD{
\left(z^2 + {3\over g}z - {T\over g} + {f\over 2}\right)^2 + P(2z) =
\left(z^2 + {3\over g}z - {T\over g} + {f\over 2}\right)^2
+ {1\over g}\left(2z-X_1\right)^2\left(2z-X_2\right) = 0
}
corresponding to the four branch points of two remaining cuts. These
cuts can be interpreted as "splittings" of the double zeroes of the
"classical limit" of this equation given by \CLAZE.  Let us denote the
splitting as $\hat z_{1,2}
\to \hat z_{1,2}^{\pm}$, so that $\hat z^\pm_{1}$ and  $\hat z^\pm_{2}$
 are four solutions to \FOUROD.  Then the only non-degenerate $A$- and
$B$- periods on the torus are given by the integrals of the
differential $dS^{2MM} = {1\over 2\pi i}\tilde zdz$ between these points
\eqn\PERDG{\eqalign{
S = {1\over 2\pi i}\oint_{\hat z^-_1}^{\hat z^+_1}\tilde {z}dz
\cr
{\p{\cal F}^{2MM}\over\p S} =
\Pi = {1\over 2\pi i}\oint_{\hat z^+_1}^{\hat z^-_2}\tilde zdz =
{1\over \pi i}\int_{\hat z^+_1}^{\hat z^-_2}\tilde zdz
}}
where $\tilde z$ is related to $z$ via
$$
\left(z\tilde z + {3\over 2g}(z+\tilde z) - {T\over g} + {f\over 2}\right)^2
+ {1\over g}\left(z+\tilde z-X_1\right)^2
\left(z+\tilde z-X_2\right) = 0.
$$
These relations for the only nontrivial period $S$ \PERDG\ on this
degenerate $g_{\rm red}=1$ curve should be supplemented by the
relations \tztmm\ and \dfttz, defining the dependence on $t_0$. As
usual, one may choose instead their linear combinations $S_1=S$ and
$S_2=t_0-S$, corresponding to the filling numbers of the two cuts.

In this way we formulated the explicit solution of the two support two
matrix model with the real cubic potential. We point here out again
that the integrals \PERDG\ can be in principle calculated in terms of
elliptic functions.

\newsec{Connection with SUSY gauge theories}

Recently it was proposed in \DV\ to build the geometries of underlying
string theories for certain ${\cal N}=1$ SUSY gauge theories by
effectively reducing them to the complex curves \dvc\ of 1MM.  The
curves of the 1MM belong to the same class as the Seiberg-Witten
curves of the $SU(n)$ ${\cal N}=2$ SUSY gauge theories with $n-1$
fundamental matter multiplets.  The understanding of this proposal
directly from the field theory was considerably advanced in \HAM, and
later in \CDSW. Though the parallels between matrix models and
four-dimensional supersymmetric gauge theories
based on the similarity of their
integrable structures were noticed much earlier \GM, the recent
observation of \DV\ contains a direct conjecture relating the
superpotentials in $\CN=1$ four dimensional theories to the partition
functions of the multi-cut solutions, like ones considered in our
paper.

According to the proposal of \DV\ the effective potentials of gaugino
condensates $S_i=\lan \Tr W^{(i)}_\a W_{(i)}^\a \ran$ in a large class
of four-dimensional $\CN=1$ gauge theories can be calculated in terms
of the planar limit of matrix integrals. For the $\CN=1$ theory with
one adjoint matter multiplet (broken $\CN=2$ supersymmetric theory)
the calculation reduces to the large $N$ solution of the one-matrix
integral \OMM, in general having multiple cuts, as described in the
previous section.  When the $U(N)$ gauge group is broken to
$U(N_1)\times U(N_2)\times\ldots\times U(N_{k})$, with the classical
VEV's of different subgroups located at the different extrema of the
tree superpotential, the matrix model predicts the values of $k$
gaugino condensates $S_i=\lan \Tr W^{(i)}_\a W_{(i)}^\a \ran$
corresponding to the the vector multiplets $W^{(i)}_\a$ of the gauge
subgroups. The effective potential $W_{eff}(S_i,\tau)$ as a function
of these condensates and the complexified gauge coupling $\tau$ can be
related to the free energy of the multi-cut solution as follows
\eqn\DVREC{
W_{eff}(S,\tau)=
\sum_i N_i \(2\pi i \tau  S_i-
{\p \CF(S_1,\ldots,S_k)\over \p S_i}\),  }
the logarithmic term being hidden in the second term representing the
derivative of the matrix model free energy,
according to the proposal of \DV.
The variables $S_1,\ldots,S_k$ appear, strictly speaking, only in the
planar limit of this matrix model and correspond to the eigenvalue
filling numbers $S_i=\hbar N_i \propto N_i/N$ of various classical
extrema of of the matrix action, giving rise to the dependence of
multi-support solutions upon the variables \DVper, \pertmm\
discussed in our paper.

For the one-matrix model \OMM\  the situation looks to be relatively simple
since the key observation comes from the fact, that from the
coincidence of the matrix model and Seiberg-Witten curves it trivially
follows that
\eqn\PEMAT{
{\p^2{\cal F}^{1MM}\over\p S_i\p S_j} = {\p^2{\cal F}^{SW}\over\p a_i\p a_j}, }
due to coincidence (after fixing the homology basis) of the period
matrices. The Seiberg-Witten prepotential ${\cal F}^{SW}$ \SW\
as a function of a {\it different} set of variables
\eqn\qcdper{
a_i =
\oint_{A_i}\lambda\left({dW'_n\over y} - {W'_n\over y}{df
\over 2f}\right)
}
is defined as
\eqn\FSW{
{\p{\cal F}^{SW}\over\p a_i} =
\oint_{B_i}\lambda\left({dW'_n\over y} - {W'_n\over y}{df\over 2f}\right),
}
and it is also a quasiclassical tau-function \GKMMM.  Note that
eq. \PEMAT\ states only that the second derivatives of different
functions in different variables coincide, but these functions
themselves are certainly different quasiclassical tau-functions. Such
identification became possible first of all since the number of
multicut variables for the 1MM is $n-1 = {\rm rank}\ SU(n)$, what is
precisely the dimension of the moduli space of (the Coulomb phase of)
the $SU(n)$ gauge theory.

For the softly broken $\CN=4$ theory, according to the proposal of
\DV, $\CF(S_1,\ldots,S_k)=\log {\cal Z}$ should be calculated as
the planar limit of the matrix integral
$$
{\cal Z}=\int \CD \Phi_1\ \CD \Phi_2\ \CD \Phi_3\
e^{\Tr\(i\Phi_1[\Phi_1,\Phi_2]+V(\Phi_1,\Phi_2,\Phi_3)\)}
$$
with $\Phi_1,\Phi_2,\Phi_3$ considered as simple hermitean $N\times N$
matrices.
Of course not all matrix integrals of this kind are calculable. An
important case corresponding to the $\CN=1^*$ perturbation
$V(\Phi_1,\Phi_2,\Phi_3)= m\sum_{i=1}^3 \Phi_i^2$ is  considered
in \DV\ and based on the planar solution of this matrix model given in
\KKN.

If only two out of three masses are nonzero and equal, the theory
possesses $\CN=2$ supersymmetry and its non-perturbative solution is
formulated in terms of the elliptic Calogero-Moser system
\DoWi,\Mart,\GM, whose spectral curve \KRICAL\ covers $n$ times the
elliptic curve. There seems yet to be no naive and direct relations
between the corresponding Seiberg-Witten theory and quasiclassical
tau-function of two matrix model, considered in this paper. However,
the structure of its complex curve suggests that certain geometric
parallels between these two theories are quite possible.

Let us discuss a possible place of the two matrix models
in the context of the proposal of \DV.
An obvious and interesting generalization of the $\CN=1$ SYM theory
with one adjoint chiral multiplet is the the case of a few multiplets
with direct interactions among the fields. If one takes, in the case
of two adjoint chiral multiplets $X$ and $Y$, the tree superpotential
$W_{tree}= -\Tr XY +\Tr W(X)+ \Tr \tilde W(Y)$ then the function
$\CF(S_1,S_2,\ldots)$ in the general formula \DVREC\ should be chosen as
the planar free energy of the two matrix model with (in general)
multiple supports, being calculated according to the equations
\pertmm\ and \dptmm.

 In this way we establish the relation
between the algebraic curve of the two matrix model and geometry of
the supersymmetric theory with the described class of tree potentials.

The two matrix models obeys a rich phase structure in the space of its
couplings. Its critical points correspond to  collisions of various
singularities on the algebraic curve, like the collapse of the B-period
of the Seiberg-Witten curve corresponding to the appearence of
 massless monopoles.
These critical points were used for the complete
classification (within the H2MM)  of the models of $(p,q)$ rational
matter fields interacting with 2D gravity  (see for example \DKK).
It would be interesting to study the consequences of this well established
picture for the phase structure of the underlying $\CN=1$ SYM theory
with two adjoint chiral multiplets.

Let us also note here that an interesting generalisation of the standard
H2MM considered in this paper is the model describing the perturbed
quantum mechanics of the inverted oscillator, proposed in \AKK.
The model
is related to the dynamics of windings in the compactified 2D string
theory. For rational $R$ it  can be also described by an
algebraic curve.

In contrast to the one matrix model, in the two matrix model the
number of multicut parameters \genus\ grows exactly as the {\it
dimension} of the $SU(n)$ group $n^2-1 = {\rm dim}\ SU(n)$.  Naively,
this would correspond to the total breaking of the $SU(n)$ gauge
group in the Seiberg-Witten like context, including even the breaking
of the corresponding global
symmetry, or to a theory with a more complicated gauge/matter
structure.

However, an important particular case of the multicut solution of the
two-matrix model corresponding to the case of real couplings in the
potential and only $n-1$ real eigenvalue supports, giving the curve
with genus \GRED.  In this case we have the number of extra parameters
exactly equal to the genus $g_{\rm red} = n-1 = {\rm rank}\ SU(n)$,
which might be more appropriate for the study of the symmetry breaking
in the corresponding $\CN=1$ SYM theory.

There exist a few obvious generalizations of the multi-matrix interactions,
giving rise to a very diverse class of the corresponding
multi-field  interactions in the $\CN=1$ SYM theories.  A rather general
class   of  interesting multi-matrix models with tree-like interactions
will be  studied in the next section.

\newsec{ Possible generalizations}

 A large class of solvable matrix models\foot{ where by
``solvability'' we mean a possibility to reduce the number of degrees
of freedom from the order of $N^2$ to the order of $N$, by integrating over the
angular variables of the matrices} can be classified by "tree
diagrams", where each edge of the tree connects two matrices sitting
at the vertices. The corresponding matrix model potential can be
written as
\eqn\VTREE{
V (\Phi) = \Tr\left(- \sum_{\rm i>j=1}^QC_{ij}\Phi_i\Phi_j +
\sum_{\rm i}W_i(\Phi_i) \right)
}
where $C_{ij}=1$, if $i,j$ are the neighboring vertices of the tree,
and $C_{ij}=0$ otherwise. A particular kind of such a solvable model
with tree-like interaction, the Potts model on planar graphs, was
first considered in \POTTS.

It is easy to integrate out the ``angular'' parts of hermitean
matrices $\Phi_i,\ \ i=1,\ldots,Q$, since they are independent in the
case of a tree interaction, taking the corresponding HCIZ integrals
and to rewrite the partition function of the model in terms of their
eigenvalues $\Phi_i= {\rm
diag}\left(z^{(i)}_1,\ldots,z^{(i)}_N\right)$ (see for example \KAZSM)
\eqn\EVTR{   Z=\int\ \prod_{k=1}^N\left(\prod_i
\left(d z^{(i)}_k e^{   W_i(z^{(i)}_k)}\right)\right)
e^{- \sum_{i,j}C_{ij}z^{(i)}_k z^{(j)}_k} \prod_{i=1}^Q
\left[\Delta(z^{(i)})\right]^{2-m_i}   }
where $m_i=\sum_{j=1}^QC_{ij}$ is the coordination number of the
$i$-th vertex.

Introducing the resolvents of matrices $G_i(z)$ (having as usual, the
assymptotics $G_i(z)\to t_0/z$ at $z\to\infty$) we can write down the
following saddle point equations, generalizing the eqs. \SPE\ of the
H2MM:
\eqn\SPET{  \sum_{j=1}^Q\ C_{ij}\ z^{(j)}-W_i'(z^{(i)}) =(m_i-2)G_i(z^{(i)})
}
As in the case of  the two matrix model, this system of equations should be
degenerate, and this degeneracy is the  condition of its solution in
terms of an algebraic hyper-surface relating all $Q$ variables\foot{ An
important paricular case of such a surface will be considered in
\CDGKV.}.  Namely, it should exist a polynomial function of $F$
depending on all  $z^{(i)}$, $i=1,\ldots,Q$, such that
\eqn\KKK{ F\left(z^{(1)},\ldots,z^{(Q)}\right)=0  }
in analogy with \POLEQ. If the system \SPET\ was not degenerate it
would give only pointlike distributions, leading to the collaps of
eigenvalues into one or a few points.

To build the function \KKK, and to analyse the structure of the
corresponding algebraic surface we should start as usual, from the
``classical'' equations (cf. with \CLASS) corresponing to putting all
the r.h.s. of \SPET\ to zero. The ``classical'' limit of the function
\KKK\ corresponds to the product of all these ``classical'' equations
(in analogy with \CUCLA\ for the two matrix model).  Then one can write:
$$
 F\left(z^{(1)},\ldots,z^{(Q)}\right)=\prod_{i=1}^Q\ \left[
 \sum_{j=1}^Q\ C_{ij}\ z^{(j)}-W_i'(z^{(i)})\right] + {\rm deformations}=0,
$$
where by the deformations we mean adding a polynomial in all variables
of lower degree, governed by corresponding multidimensional Newton
polyhedron. The coefficients in front of the monomials of higher
degrees are determined by the assymptotics at infinities following
from \SPET\ and coincide with their values in the classical part. The
rest of the deformation coefficients will provide the new moduli of
the complex structure of this algebraic manifold.

The algebraic equation \KKK\ has the degree ${\it deg}=\prod_{i=1}^Q
(K_i-1)$, where $K_i$ is the highest power of the potential
$W_i(z)$. This corresponds to the number of extrema in the classical
multi-matrix  potential and to the number of moduli parameters of the curve
contained in the deformation.

The fact that this algebraic surface is in general not a curve does
not contradict the existence of the resolvents $G_i(z)$, which means that
 for any two variables $z^{(i)},z^{(j)}$ it should exist an algebraic
curve $F_{ij}(z^{(i)},z^{(j)})=0$ with
a polynomial function $F_{ij}(x,y)$ of the same  degree.
This loss of information in the abovementioned surface
might be related to the existence of higher dimensional holomorphic
differential forms, like the well known three-form $d\Omega_3$ on the
Calabi-Yau 3-folds. On particular 3-cycles, after the integration over
two variables it turns into a meromorphic one-differential.  This
differential is related to the resolvent with respect to the third
variable \VafaR. In principle, we could try to restore the individual
algebraic curves out of
the general surface by excluding the variables one by one from the
equation for the surface, using   \SPET.

Of course it is only a sketch of the construction of the algebraic
curve of the tree-like multi-matrix model. It would be interesing to
precise the details and the structure of these  algebraic curves,
although it might be difficult to do it to the same extent of
explicitness as we have done in this paper for the two matrix model.

Some further generalisations are possible if we substitute the
potential \VTREE\ by
\eqn\VTREE{
V (\Phi) = \Tr\left(- \sum_{\rm i>j=1}^QC_{ij}(\Phi_i\Phi_j) +
\sum_{\rm i}W_i(\Phi_i) \right)
}
where $C_{ij}(M)$ are arbitrary polynomial functions (nonzero only on
a tree). In this case we can achieve the reduction to the eigenvalues
by the method of character expansion (see \KAZSM\ and references
therein). Namely, we can expand the exponent of each interaction term
into the $GL(N)$ characters $\chi_R(M)$
$$
e^{-C_{ij}\Tr (\Phi_i\Phi_j)}=\sum_R f^{(i,j)}_R \chi_R(\Phi_i\Phi_j)
$$
and then use the orthogonality property of the matrix elements to
integrate out the relative angle of two matrices. An example of such
calculation was done in \KZJ\ for the $\Tr(\Phi_i\Phi_j)^2$
interaction.  The corresponding system of polynomial equations will
include, along with the eigenvalue variables, the dual variables
corresponding to the highest weights of the Young tableaux of the
$SL(N)$ irreducible representations. Here the construction of the
algebraic surface looks even more difficult, but certainly not
impossible. Note that the model still stays ``solvable'' by the
character expansion method if we change the arguments of products
$C_{ij}(\Phi_i\Phi_j)$ by $C_{ij}(\Phi_i^{k_{ij}}\Phi_j^{n_{ij}})$
with arbitrary integers $k_{ij},n_{ij}$ for each $ij$-link.

These models (and those which can be reduced to them by introducing
some gaussian matrix integrations, like in the case of the Potts model
on planar graphs \POTTS) exhaust the list of ``solvable''
(i.e., reducible to the eigenvalues) multi-matrix models.

\newsec{Conclusion}

In this paper we studied the multi-support solutions of two matrix
models and we found that they lead to the appearence of a new nontrivial
one-dimensional complex geometry. The corresponding quasiclassical
tau-function can be still defined in a standard way and even rather
explicit formulas defining the tau-funcion can be written down.

The main problem for the multi-support solutions is nevertheless to
write down explicitly the system of integrable equations which is
solved by the corresponding free energy. For the one-support solutions
such system necessarily includes the dispersionless Hirota equations
which have a nice and well-known dispersionfull analog.

The only known analogs of the dispersionless Hirota equations for the
multi-support case are the associativity or WDVV equations \WDVV, and
a wide class of their solutions is constructed on the base of
quasiclassical tau-functions.  The formula \ZERDZ\ in particular
suggests that the quasiclassical partition function of the cubic two
matrix model in the "symmetric ansatz", say with $t_0=0$, satisfies
the WDVV equations \WDVV,\MMM. This follows from simple counting of
the number of variables (six coefficients of the potential and three
filling numbers altogether give the same number of free parameters as
the number of critical points \ZERDZ\ for $n=2$) and the structure of
the residue formula for this case. We mention this fact since it seems
to be the only explicitly known solution to WDVV equations coming from
the nonhyperelliptic curves (see \MaWDVV\ for a general discussion of
this issue).

In this paper we presented a two-support solution of the two matrix
model for the cubic potential in a rather explicit form. Nevertheless,
in principle it can be precised further. In particular, the period
integrals \PERDG\ are elliptic (according to the structure of the
underlying curve having the genus $g_{\rm red}=1$) and can be in
principle calculated explicitly. It might be instructive to do it and
to write down explicitly their asymptotic expansion.

The multi-solutions we found are quite interesting from the point of
view of  statistical-mechanical models on planar graphs. An example
of such model (a double-phase Ising model) is described in section 2.
It would be interesting to classify all such possible models emerging
from the multi-support two matrix model.

From the point of view of the underlying $\CN=1$ SYM theory it is very
desirable to study the degenerations of a higher genus algebraic curve
(with more than two cuts filled)  of the two matrix model considered
above and classify the emerging physical excitations (monopoles, dyons
etc), by analogy to the hyperelliptic solution in (generalized)
Seiberg-Witten picture.

Finally, the models described in section 7 should contain a much
richer variety of possible algebraic surfaces describing their planar
limit. They certainly deserve a considerable attention.
The matrix models, due to their natural integrability
properties, could give an insight into the structure of algebraic
surfaces possessing interesting physical applications.

\bigskip
\noindent{\bf Acknowledgements:}

We are indebted to M.Aganagic, S.Alexandrov,
E.Brezin, A.Gorsky, S.Gukov, H.Itoyama, I.Kostov,
A.Levin, A.Mironov, A.Morozov, N.Nekrasov, V.Rubtsov and P.Wiegmann for
many interesting discussions and especially to R.Dijkgraaf, I.Krichever,
P.Push\-kar and A.Zabrodin for very helpful conversations. The work of
V.K. was partially
supported by European Union under the RTN contracts HPRN-CT-2000-00122
and -00131. The work of
A.M. was also partially supported by RFBR grant 02-02-16496, INTAS
grant 00-00334 and grant of support of scientific schools
No.~00-15-96566.
A.M. would like also to thank Laboratoire de Physique
Th\'eorique de l'\'Ecole Normale Sup\'erieure and Institut des Hautes
\'Etudes Scientifiques for the warm hospitality.

\listrefs
\bye